\renewcommand{\Re}{\mathop{\rm Re}\nolimits}
\def\slash#1{\setbox0=\hbox{$#1$}               
   \dimen0=\wd0                                 
   \setbox1=\hbox{/} \dimen1=\wd1               
   \ifdim\dimen0>\dimen1                        
      \rlap{\hbox to \dimen0{\hfil/\hfil}}      
      #1                                        
   \else                                        
      \rlap{\hbox to \dimen1{\hfil$#1$\hfil}}   
      /                                         
   \fi}                                         %

\documentclass[12pt]{article}
\usepackage{epsfig}
\usepackage{cite}
\usepackage{amsmath}

\begin{document}
\begin{titlepage}
\begin{flushright}
UM-TH-97-17\\
May 1998\\
\end{flushright}
\vskip 2cm
\begin{center}
{\large\bf ${\cal O}(N_f\alpha^2)$ Electromagnetic Charge Renormalization\\
           in the Standard Model}
\vskip 1cm
{\large Paresh Malde and  Robin G. Stuart}
\vskip 1cm
{\it Randall Laboratory of Physics,
 University of Michigan,\\
 Ann Arbor, MI 48109-1120,\\
 USA\\}
\end{center}
\vskip .5cm
\begin{abstract}
The ${\cal O}(N_f\alpha^2)$ electroweak radiative
corrections to the Thomson
scattering matrix element are calculated for a general renormalization
scheme with massless fermions.
All integrals can be evaluated exactly in dimensional
regularization which in several cases yields new results that are
remarkably simple in form.
A number of stringent internal consistency checks
are performed. The $Z$-$\gamma$ mixing complicates the calculation of
1-particle reducible diagrams considerably
at this order and a general treatment of this problem is given.
Conditions satisfied by ${\cal O}(N_f\alpha^2)$
counterterms are derived and may be applied to other calculations at this
order.
\end{abstract}
\vskip 4cm
\end{titlepage}

\setcounter{footnote}{0}
\setcounter{page}{2}
\setcounter{section}{0}
\newpage

\section{Introduction}

A century ago the electron was discovered by J.\ J.\ Thomson. Today the
scattering process that bears his name serves as a means to define the
most precisely measured physical constant, the electromagnetic coupling
constant, $\alpha$. Thomson scattering is the scattering of a fermion, most
conveniently taken to be an electron, off a single photon of vanishingly
small energy. It is only for such a photon that both energy and momentum
for the process
can be simultaneously conserved. The cross-section for Thomson scattering
is
\begin{equation}
\sigma_T=\frac{8\pi\alpha}{3m_f^2}Q_f^2,
\end{equation}
where $m_f$ is the fermion's mass and $Q_f$ is its charge. It thus
provides direct access to the strength of the electromagnetic coupling,
$\alpha$. Nowadays $\alpha$ is determined from the quantum Hall
effect or from the electron's anomalous magnetic moment,
but Thomson scattering remains the
prototypical process by which $\alpha$ is defined in particle physics.

A general renormalizable model contains a number of free parameters that
must
be fixed by experimental input in order for the model to become predictive.
For highest precision the best measured quantities are used. In the case
of the Standard Model of electroweak interactions there are three free
parameters of the bosonic sector that are normally fixed using
$\alpha$, $G_\mu$, the muon decay constant, and $M_Z$ the mass of the
$Z^0$ boson. Of these $\alpha$ is by far the best known quantity
experimentally. Its use in high-energy calculations introduces
contributions from hadronic effects that can be mitigated to some
extent using dispersion relations applied to experimental data
\cite{Zeppenfeld,Jegerlehner,Swartz,DavierHocker,KuhnSteinhauser}.
Still an hadronic uncertainty
remains but this may be eliminated using one extra piece of experimental
data as input \cite{StuartHadron}. $\alpha$ then plays a key r\^ole
in all calculations of precision electroweak physics.

In recent years great strides have been made in the calculation of
2-loop and higher-order Feynman diagrams and their application to
electroweak physics. In the processes considered to date, the main
effort has been directed at the computation of the Feynman
diagrams whereas the renormalization has normally been a very
minor aspect. In the calculation of the
${\cal O}(\alpha^2 m_t^4/M_W^2)$ corrections to the $\rho$-parameter
\cite{Barbieri1,Barbieri2,Fleischer1,Fleischer2}
there is a single topology containing counterterms. In some calculations
\cite{DegrGambVici} the $\overline{\rm MS}$ renormalization scheme is used
and the counterterms are taken care of by discarding divergent pieces
of diagrams. Such an approach is dangerous as it removes the
check of the cancellation of divergences between diagrams and counterterms.
The calculations are then performed in more than one gauge to test for
consistency. The full ${\cal O}(\alpha^2)$ corrections to the anomalous
magnetic moment of the muon have been calculated \cite{CKM1,CKM2}.
This process first appears at ${\cal O}(\alpha)$ and so ${\cal O}(\alpha^2)$
counterterms are not encountered and much of the complexity associated
with 2-loop renormalization in the Standard Model is avoided.

As progress continues in 2-loop calculations, confrontation with the full
complexity of 2-loop renormalization is inevitable. It is our aim here
to calculate the 2-loop corrections to Thomson scattering in a general
renormalization scheme. We will limit ourselves to corrections that
contain an internal fermion loop with the fermions assumed to be massless.
These will be referred to as ${\cal O}(N_f\alpha^2)$ corrections where
$N_f$ is the number of fermions. Because $N_f$ uniquely tags these
corrections, they form a separately gauge-invariant set and can be
expected to be dominant because $N_f$ is quite large. Although the
${\cal O}(N_f\alpha^2)$ corrections are a somewhat reduced set, the full
complexity of the 2-loop calculation is manifest and all the intricacies
and new features of the 2-loop renormalization are present. The calculation
will allow us to obtain conditions on the counterterms which, once
obtained, can be used in other calculations of this order. The resulting
cancellation of divergences provides a powerful check that is an
alternative to calculating in several gauges.

In the present calculation, one of the greatest challenges is
organizational. The various contributions must be arranged
so as to avoid double
counting and should be grouped in some logically consistent fashion.
Decisions have to be made as to whether to carry wavefunction
counterterms in individual diagrams, and thereby work with finite
Greens functions, even when they can be shown to cancel in the final
matrix element. The $Z$-$\gamma$ mixing adds considerably to the complexity
of the calculation at this order. The presentation chosen
here is an attempt to achieve the goals of consistency and clarity.
Contributions that are
clearly related, such as the 2-loop $Z$-$\gamma$ mixing and photon vertex
corrections are treated together as far as possible.

In section~2 our notation is explained. Section~3 explains the
renormalization of the Standard model and derives the relevant
counterterms valid at 2-loops. Section~4 reviews the calculation of
Thomson scattering at ${\cal O}(\alpha)$. Section~5 discusses the r\^ole
of wavefunction renormalization and Ward identities in the calculation.
It is shown that the 1-loop Ward identities must be explicitly imposed.
In section~6 the calculation of Thomson scattering at
${\cal O}(N_f\alpha^2)$ is described with separate subsections devoted
to the various classes of contributions. Finally section~7 derives
conditions that must be satisfied by the 2-loop counterterms.

\section{Notation and Conventions}

In calculating radiative corrections
to ${\cal O}(N_f\alpha^2)$, expressions will be obtained
that contain the product of two 1-loop contributions and it will thus
be necessary to distinguish between 1-loop fermionic and bosonic
corrections. The order and type of a correction will be indicated,
where needed, by a superscript in parentheses. Thus $\delta Z^{(1f)}$
indicates the 1-loop fermionic part of the counterterm $\delta Z$.
The 1-loop bosonic corrections are denoted by the superscript ${}^{(1b)}$
and the superscript ${}^{(1)}$ indicates both together. The superscript
${}^{(2)}$ when used here means the full ${\cal O}(N_f\alpha^2)$ correction.

Ultraviolet (UV) divergences will be regulated by dimensional
regularization in which $n$ denotes the complex number of space-time
dimensions. Most loop integrals will be given in exact form rather than
expanding about $n=4$. In fact, keeping the full $n$ dependence helps
display some dramatic cancellations that occur between different
classes of Feynman diagrams. It is assumed that nowadays, with the
wide availability of computer algebraic manipulation programs, the
exact results are easily transformed into series expansions when
required. With this in mind some expressions are conveniently
and compactly written in terms of $\epsilon=2-n/2$.

Two-point functions for the vector bosons, $\Pi_{\mu\nu}(q^2)$, can
always be divided into transverse and longitudinal pieces,
\[
\Pi_{\mu\nu}(q^2)=\left(\delta_{\mu\nu}-\frac{q_\mu q_\nu}{q^2}\right)
                  \Pi_T(q^2)
                 +\left(\frac{q_\mu q_\nu}{q^2}\right)\Pi_L(q^2).
\]
Here we will be exclusively concerned with the transverse part $\Pi_T(q^2)$.
The subscript `${}_T$' will therefore be dropped.

Throughout this work the Euclidean metric is used with the square of
time-like being negative. The calculation is performed in
't~Hooft-Feynman, $R_{\xi=1}$, gauge.

A fully anti-commuting Dirac $\gamma_5$ will be assumed. This could only
lead to difficulties in fermion loops that generate the antisymmetric
$\epsilon$ tensor, such as internal fermion triangles. In that case,
however, when one sums over a complete generation, anomaly cancellation
guarantees that additional terms cannot appear.

\section{Renormalization of the Standard Model}

\subsection{Renormalization of the Bosonic Sector}

The bare lagrangian, ${\cal L}^0$ is the true lagrangian of the theory.
The renormalized lagrangian, ${\cal L}^R$ and counterterm lagrangian,
$\delta{\cal L}$, satisfy ${\cal L}^0={\cal L}^R+\delta{\cal L}$.
The bare lagrangian of the Standard Model for free neutral gauge bosons
is
\begin{equation}
{\cal L}_W^0=-\frac{1}{2}(\partial_\nu W^0_{3\mu})\partial_\nu W^0_{3\mu}
             -\frac{1}{2}(\partial_\nu B^0_\mu)\partial_\nu B^0_\mu
           -\frac{(M_W^2)^0}{2{g^0}^2}
           \left(g^{\prime0}B_\mu^0-g^0W_{3\mu}\right)^2
\label{eq:freelagrangian}
\end{equation}
where $W^0$ and $B^0$ are the bare $SU(2)_L$ isospin and $U(1)$
hypercharge fields
and $g^0$ and $g^{\prime0}$ are the $SU(2)_L$ and $U(1)$ coupling
constants respectively.
$(M_W^2)^0$ is the bare mass squared of the
$W$ boson and $(M_Z^2)^0$ will be used to denote that of the $Z^0$
boson. They are constructed from parameters of the Higgs sector
from which the relation
\begin{equation}
\frac{(M_W^2)^0}{(M_Z^2)^0}=\frac{{g^0}^2}{{g^0}^2+{g^{\prime0}}^2}
\label{eq:baremassratio}
\end{equation}
may be derived.
The renormalized and counterterm lagrangians are obtained by writing
the bare fields, coupling constants and masses in terms of the
corresponding renormalized quantities and their counterterms,
\begin{alignat}{3}
W^0&=(1+\delta Z_W)^{\frac{1}{2}}W&\qquad\qquad
g^0&=g+\delta g&\qquad\qquad
(M_W^2)^0&=M_W^2+\delta M_W^2\label{eq:renorm1}\\
B^0&=(1+\delta Z_B)^{\frac{1}{2}}B&\qquad\qquad
g^{\prime0}&=g^\prime+\delta g^\prime&\qquad\qquad
(M_Z^2)^0&=M_Z^2+\delta M_Z^2\label{eq:renorm2}
\end{alignat}
These conventions differ from those used Ross and Taylor
\cite{RossTaylor} and are more closely consistent with those of
Aoki {\it et al} \cite{Aokietal}.

The weak mixing angle, $\theta_W$, is defined so as to diagonalize the
mass matrix of the renormalized fields and hence the renormalized
$W_3$ and $B$ fields are then related to renormalized $Z$ field
and photon field, $A$, by
\begin{alignat}{2}
W_3&=&c_\theta Z+s_\theta A,\\
B&=-&s_\theta Z+c_\theta A,
\end{alignat}
where $s_\theta$ and $c_\theta$ are $\sin\theta_W$ and $\cos\theta_W$
respectively.

The transverse parts of the 2-point counterterms for the photon,
$Z^0$ and $Z$-$\gamma$ mixing
are obtained by substituting eq.s(\ref{eq:renorm1}) and
(\ref{eq:renorm2}) into (\ref{eq:freelagrangian}).
Keeping only terms that can contribute up second order at $q^2=0$ gives

\begin{eqnarray}
\raisebox{-0.2cm}{
\begin{minipage}[t][0.5cm][c]{2.5cm}
\epsfig{file=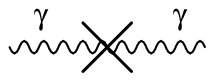,width=2.5cm}
\end{minipage}}
&=&-\frac{1}{2}M_Z^2(2s_\theta c_\theta\delta_-
     +\frac{1}{2}\delta Z_{Z\gamma})^2
\label{eq:photon2ptct}\\
\raisebox{-0.2cm}{
\begin{minipage}[t][0.5cm][c]{2.5cm}
\epsfig{file=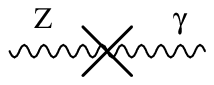,width=2.5cm}
\end{minipage}}
&=&-M_Z^2\bigg\{(2s_\theta c_\theta\delta_-
                   +\frac{1}{2}\delta Z_{Z\gamma})
\nonumber\\ & & \ \ \ \ \ \ \
                   -\frac{(s_\theta^2-c_\theta^2)}{8s_\theta c_\theta}
                   \delta Z_{Z\gamma}^2
                  +\left(\frac{\delta M_W^2}{M_W^2}
                         -2s_\theta^2\delta_-\right)
                        (2s_\theta c_\theta\delta_-
                       +\frac{1}{2}\delta Z_{Z\gamma})
\nonumber\\ & & \ \ \ \ \ \ \
                  +s_\theta c_\theta\delta_-(\delta Z_B+\delta Z_W)
                  -2s_\theta c_\theta\delta_-\frac{\delta g}{g}\bigg\}
\label{eq:Zgamma2ptct}\\
\raisebox{-0.2cm}{
\begin{minipage}[t][0.5cm][c]{2.5cm}
\epsfig{file=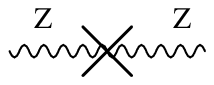,width=2.5cm}
\end{minipage}}
&=&-\delta M_Z^2-M_Z^2\delta Z_Z\nonumber
\\ & &-M_Z^2\bigg\{\frac{1}{4}\delta Z_{Z\gamma}^2
                 +2\frac{s_\theta}{c_\theta}
                  \frac{\delta M_W^2}{M_W^2}
                       (2s_\theta c_\theta\delta_-
                       +\frac{1}{2}\delta Z_{Z\gamma})
                    -\frac{\delta M_W^2}{M_W^2}\delta Z_W
\nonumber\\ & & \ \ \ \ \ \ \
                  -4s_\theta^4\delta_-^2
                  -2\frac{s_\theta^3}{c_\theta}\delta_-\delta Z_{Z\gamma}
                  +2\frac{s_\theta}{c_\theta}\delta_-(\delta Z_B+\delta Z_W)
                  -4s_\theta^2\delta_-\frac{\delta g}{g}\bigg\}
\nonumber\\
\end{eqnarray}
where here and in what follows we define
\begin{gather}
\delta Z_\gamma=c_\theta^2\delta Z_B+s_\theta^2\delta Z_W,\quad
\delta Z_Z=s_\theta^2\delta Z_B+c_\theta^2\delta Z_W,\quad
\delta Z_{Z\gamma}=s_\theta c_\theta(\delta Z_W-\delta Z_B),\\
\delta_-=\frac{1}{2}\left(\frac{\delta g}{g}
                         -\frac{\delta g^\prime}{g^\prime}\right).
\end{gather}
The relation
\begin{equation}
 \frac{\delta M_W^{2(1)}}{M_W^2}
-\frac{\delta M_Z^{2(1)}}{M_Z^2}=4s_\theta^2\delta_-^{(1)}
\end{equation}
follows from eq.(\ref{eq:baremassratio}) and is valid to first
order. Note that at this order
the photon develops a mass counterterm \cite{BaulieuCoquereaux}
which is a
reflection of the fact that the renormalized field, $A_\mu$, is not the
same as the physical photon. In principle one could rediagonalize the
neutral boson mass matrix at each order by redefining $\theta_W$ but
it is cumbersome and unnecessary.

In the corrections to Thomson scattering the charged $W$-boson
appears as an internal particle. Its ${\cal O}(\alpha)$ 2-point
counterterm is
\begin{equation}
\raisebox{-0.2cm}{
\begin{minipage}[t][0.5cm][c]{2.5cm}
\epsfig{file=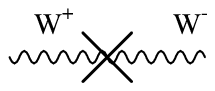,width=2.5cm}
\end{minipage}}
=-\delta Z_W^{(1)}(q^2+M_W^2)\delta_{\mu\nu}
 +\delta Z_W^{(1)} q_\mu q_\nu
 -\delta M_W^{2(1)}\delta_{\mu\nu}
\label{eq:W2ptct}
\end{equation}

Note that the first term in eq.(\ref{eq:W2ptct}) contains an inverse
propagator, $(q^2+M_W^2)$, and so for internal $W$'s it will cancel with
wavefunction counterterms at vertices. Of course coupling constant
counterterms and external particle wavefunction counterterms still
need to be included but internal $W$'s therefore effectively only
generate 2-point counterterms of the form
$\delta Z_W^{(1)} q_\mu q_\nu-\delta M_W^{2(1)}\delta_{\mu\nu}$.
In practice this provides a very
convenient way of eliminating much of the labour in calculating diagrams
constructed by inserting ${\cal O}(N_f\alpha)$ counterterms in
${\cal O}(\alpha)$ diagrams.

In the on-shell renormalization scheme the 1-loop counterterms are
\begin{eqnarray}
\delta{M_W^{2(1)}}&=&\Re\Pi_{WW}^{(1)}(-M_W^2),\label{eq:dMW}\\
\delta{M_Z^{2(1)}}&=&\Re\Pi_{ZZ}^{(1)}(-M_Z^2),\label{eq:dMZ}\\
\frac{\delta g^{(1)}}{g}&=&-\frac{1}{2}\Pi_{\gamma\gamma}^{(1)\prime}(0)
            +\frac{s_\theta}{c_\theta}\frac{\Pi_{Z\gamma}^{(1)}(0)}{M_Z^2}
            +\frac{c_\theta^2}{2s_\theta^2}
                  \Re\left(\frac{\Pi_{WW}^{(1)}(-M_W^2)}{M_W^2}
                          -\frac{\Pi_{ZZ}^{(1)}(-M_Z^2)}{M_Z^2}\right),
\ \ \ \label{eq:dg}\\
\frac{\delta g^{\prime(1)}}{g^\prime}&
         =&-\frac{1}{2}\Pi_{\gamma\gamma}^{(1)\prime}(0)
           +\frac{s_\theta}{c_\theta}\frac{\Pi_{Z\gamma}^{(1)}(0)}{M_Z^2}
           -\frac{1}{2}\Re\left(\frac{\Pi_{WW}^{(1)}(-M_W^2)}{M_W^2}
                          -\frac{\Pi_{ZZ}^{(1)}(-M_Z^2)}{M_Z^2}\right).
\label{eq:dgp}
\end{eqnarray}
In the $\overline{\rm MS}$ renormalization scheme the counterterms
are just the divergent parts of these plus certain other constants.
By direct calculation of the diagrams concerned the divergent parts of
the 1-loop counterterms are found to be
\begin{eqnarray}
\delta{M_W^2}^{(1b)}&=&\left(\frac{g^2}{16\pi^2}\right)
    \frac{M_W^2}{6c_\theta^2}(31s_\theta^2-25)\Delta\\
\delta{M_W^2}^{(1f)}&=&\left(\frac{g^2}{16\pi^2}\right)
    \frac{4M_W^2}{3}\Delta\\
\delta{M_Z^2}^{(1b)}&=&-\left(\frac{g^2}{16\pi^2}\right)
    \frac{M_Z^2}{6c_\theta^2}(42s_\theta^4-74s_\theta^2+25)\Delta\\
\delta{M_Z^2}^{(1f)}&=&\left(\frac{g^2}{16\pi^2}\right)
    \frac{4M_Z^2}{9c_\theta^2}(8s_\theta^4-6s_\theta^2+3)\Delta\\
\delta Z_W^{(1b)}&=&\left(\frac{g^2}{16\pi^2}\right)\frac{19}{6}\Delta\\
\delta Z_W^{(1f)}&=&-2\frac{\delta g^{(1f)}}{g}
   \ =\ -\left(\frac{g^2}{16\pi^2}\right)\frac{4}{3}\Delta\\
\delta Z_B^{(1b)}&=&-2\frac{\delta g^{\prime(1b)}}{g^\prime}
   \ =\
-\left(\frac{g^2}{16\pi^2}\right)\frac{s_\theta^2}{6c_\theta^2}\Delta\\
\delta Z_B^{(1f)}&=&-2\frac{\delta g^{\prime(1f)}}{g^\prime}
   \ =\
-\left(\frac{g^2}{16\pi^2}\right)\frac{20s_\theta^2}{9c_\theta^2}\Delta\\
\frac{\delta g^{(1b)}}{g}
   &=&-\left(\frac{g^2}{16\pi^2}\right)\frac{43}{12}\Delta
\end{eqnarray}
in which $\Delta=\pi^{-\epsilon}\Gamma(\epsilon)$.
In all cases the fermionic counterterms have been summed over a single
complete generation.

\subsection{Renormalization of the Fermionic Sector}

\subsubsection{Fermion 2-point counterterm}

The bare lagrangian for a free fermion, $f$, is given in terms of the bare
fermion field, $\psi^0$, and fermion mass, $m^0_f$ by
\begin{equation}
{\cal L}_\psi=-\bar\psi^0(\slash\partial+m_f^0)\psi^0.
\end{equation}
In the Standard Model the left- and right-helicity, $\psi_L$ and $\psi_R$,
components of the bare field are renormalized independently.
The renormalized fields and fermion mass are defined from
the corresponding bare quantities by the rescalings
\begin{equation}
\psi^0_L=(1+\delta Z_L)^{\frac{1}{2}}\psi_L,\ \ \ \ \ \ \
\psi^0_R=(1+\delta Z_R)^{\frac{1}{2}}\psi_R,\ \ \ \ \ \ \
m_f^0=m_f+\delta m_f.
\end{equation}
This leads to a 1-loop counterterm
\begin{eqnarray}
\raisebox{-0.5cm}{
\begin{minipage}[t][0.5cm][c]{2.5cm}
\epsfig{file=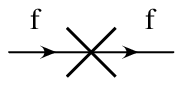,width=2.5cm}
\end{minipage}}
&=&-i\slash p (\delta Z_L\gamma_L+\delta Z_R\gamma_R)
     -m_f\left(\frac{\delta Z_L+\delta Z_R}{2}\right)
     -\delta m_f\\
  &=&-\frac{1}{2}(\delta Z_R\gamma_L+\delta Z_L\gamma_R)(i\slash p+m_f)
\nonumber\\ & &
     -\frac{1}{2}(i\slash p+m_f)(\delta Z_L\gamma_L+\delta Z_R\gamma_R)
     -\delta m_f
\label{eq:inversepropform}
\end{eqnarray}
where the form (\ref{eq:inversepropform}) is expressed in terms of
inverse propagators and is useful for demonstrating the cancellation
of fermion wavefunction counterterms where they occur in internal loops.

\begin{figure}[t]
\centerline{\epsfig{file=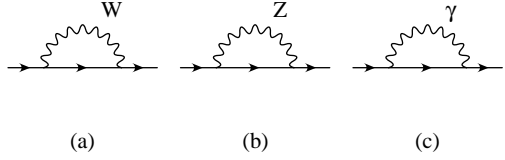}}
\caption{Fermion self-energy diagrams}
\label{fig:fermionse}
\end{figure}

The finite pieces of the counterterms will depend on the particular
renormalization scheme that has been chosen but the divergent part
is common to all schemes. This can be found by computing the 1-loop
Feynman diagrams contributing to the fermion self-energy. These diagrams
are shown in Fig.\ref{fig:fermionse} and give
\begin{eqnarray}
\raisebox{-0.25cm}{
\begin{minipage}[t][0.5cm][c]{2.5cm}
\epsfig{file=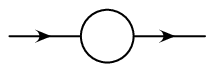,width=2.5cm}
\end{minipage}}
&=&-i\slash p\left(\frac{g^2}{16\pi^2}\right)
     \left(\frac{1}{2}\right)\gamma_L
     (\pi M_W^2)^{\frac{n}{2}-2}
     \frac{2(n-2)}{n}\Gamma\left(2-\frac{\displaystyle n}{\displaystyle
2}\right)
     \nonumber\\
  & &-i\slash p\left(\frac{g^2}{16\pi^2}\right)
     (\beta_{Lf}^2\gamma_L+\beta_{Rf}^2\gamma_R)
     (\pi M_Z^2)^{\frac{n}{2}-2}
     \frac{2(n-2)}{n}\Gamma\left(2-\frac{\displaystyle n}{\displaystyle
2}\right)
     \nonumber\\
  & &\ \ \ \ \ \ \ \ \ \ \ \ -m_f\left(\frac{g^2}{16\pi^2}\right)
     \beta_{Lf}\beta_{Rf}
     (\pi M_Z^2)^{\frac{n}{2}-2}
     \frac{2n}{(n-2)}\Gamma\left(2-\frac{\displaystyle n}{\displaystyle
2}\right)
     \nonumber\\
  & &-i\slash p\left(\frac{g^2 s_\theta^2}{16\pi^2}\right)Q_f^2
     (\pi m_f)^{\frac{n}{2}-2}
     \frac{1}{(n-3)}\Gamma\left(2-\frac{\displaystyle n}{\displaystyle
2}\right)
     \nonumber\\
  & &\ \ \ \ \ \ \ \ \ \ \ \ -m_f\left(\frac{g^2
s_\theta^2}{16\pi^2}\right)Q_f^2
     (\pi m_f)^{\frac{n}{2}-2}
     \frac{n}{(n-3)}\Gamma\left(2-\frac{\displaystyle n}{\displaystyle
2}\right)
\label{eq:fermionselfenergy}
\end{eqnarray}
where terms that are suppressed by factors $m_f^2/M_W^2$ relative to
the leading ones have been dropped and $\beta_{Lf}$ and $\beta_{Rf}$
are the left- and right-handed couplings of the $Z^0$ to the fermion,
\begin{equation}
\beta_{Lf}=\frac{t_{3f}-s_\theta^2 Q_f}{c_\theta},\ \ \ \ \ \
\beta_{Rf}=-\frac{s_\theta^2 Q_f}{c_\theta}.
\end{equation}
The first term of eq.(\ref{eq:fermionselfenergy}) comes from
Fig.\ref{fig:fermionse}a,
the second and third terms from Fig.\ref{fig:fermionse}b
and the last two from the pure QED diagram, Fig.\ref{fig:fermionse}c.

In the on-shell renormalization scheme, defined by setting the renormalized
fermion mass $m_f$ equal to the pole mass, the fermion mass counterterm
is then given by
\begin{eqnarray}
\delta m_f &=&m_f\left(\frac{g^2}{16\pi^2}\right)\left(\frac{1}{4}\right)
     (\pi M_W^2)^{\frac{n}{2}-2}
     \frac{2(n-2)}{n}\Gamma\left(2-\frac{\displaystyle n}{\displaystyle
2}\right)
     \nonumber\\
  &+&m_f\left(\frac{g^2}{16\pi^2}\right)
     \left(\frac{\beta_{Lf}^2+\beta_{Rf}^2}{2}\right)
     (\pi M_Z^2)^{\frac{n}{2}-2}
     \frac{2(n-2)}{n}\Gamma\left(2-\frac{\displaystyle n}{\displaystyle
2}\right)
     \nonumber\\
  &-&m_f\left(\frac{g^2}{16\pi^2}\right)
     \beta_{Lf}\beta_{Rf}
     (\pi M_Z^2)^{\frac{n}{2}-2}
     \frac{2n}{(n-2)}\Gamma\left(2-\frac{\displaystyle n}{\displaystyle
2}\right)
     \nonumber\\
  &-&m_f\left(\frac{g^2}{16\pi^2}\right)Q_f^2s_\theta^2
     (\pi m_f)^{\frac{n}{2}-2}
     \frac{(n-1)}{(n-3)}\Gamma\left(2-\frac{\displaystyle n}{\displaystyle
2}\right)
\label{eq:fermionmassct}
\end{eqnarray}

\subsection{Fermion-Boson interaction lagrangian}

The bare interaction lagrangian between the neutral gauge
bosons and fermions is
\begin{equation}
{\cal L}_{\psi W}^0=ig^0t_3\bar\psi^0_L\gamma_\mu\psi^0_L W^0_{3\mu}
    +ig^{\prime 0}\frac{Y_L}{2}\bar\psi^0_L\gamma_\mu\psi^0_L B^0_\mu
    +ig^{\prime 0}\frac{Y_R}{2}\bar\psi^0_R\gamma_\mu\psi^0_R B^0_\mu.
\label{eq:interactlagrangian}
\end{equation}
$\psi$ represents the fermion wavefunction.
Here $\gamma_\mu$ are the usual Dirac $\gamma$-matrices.
The electric charge, $Q$, of given fermion flavour and helicity is
related to its hypercharge, $Y$, by $Q=t_3+Y_L=Y_R$.

Substituting eq.s(\ref{eq:renorm1}) and
(\ref{eq:renorm2}) into (\ref{eq:interactlagrangian}) yields the
vertex counterterms for the photon. To second order in the
counterterms for the coupling constants and bosonic fields
and first order in the fermionic counterterms
this is
\begin{align}
\raisebox{-0.2cm}{
\begin{minipage}[t][0.5cm][c]{2.5cm}
\epsfig{file=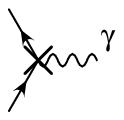,width=2.5cm}
\end{minipage}}
=igs_\theta\gamma_\mu\gamma_L\bigg\{&Q\delta Z_L
      +t_3\left(\frac{\delta g}{g}+\frac{1}{2}\delta Z_W\right)
       +(Q-t_3)\left(\frac{\delta g^\prime}{g^\prime}
                    +\frac{1}{2}\delta Z_B\right)\notag\\
      -&\frac{t_3}{2}\delta Z_W\left(\frac{1}{4}\delta Z_W
                               -\frac{\delta g}{g}\right)
      -\frac{(Q-t_3)}{2}\delta Z_B\left(\frac{1}{4}\delta Z_B
                             -\frac{\delta g^\prime}{g^\prime}\right)
      \bigg\}\notag\\
  +igQs_\theta\gamma_\mu\gamma_R\bigg\{&\delta Z_R
       +\left(\frac{\delta g^\prime}{g^\prime}
              +\frac{1}{2}\delta Z_B\right)
       -\frac{1}{2}\delta Z_B\left(\frac{1}{4}\delta Z_B
                             -\frac{\delta g^\prime}{g^\prime}\right)
      \bigg\}
\label{eq:photonvertexct}
\end{align}
and for the $Z^0$, although it is not required here, the
vertex counterterm is
\begin{align}
\raisebox{-0.2cm}{
\begin{minipage}[t][0.5cm][c]{2.5cm}
\epsfig{file=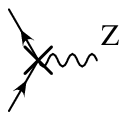,width=2.5cm}
\end{minipage}}
=ig\gamma_\mu\gamma_L\bigg\{&\beta_{L}\delta Z_L
      +t_3c_\theta\left(\frac{\delta g}{g}+\frac{1}{2}\delta Z_W\right)
       -(Q-t_3)\frac{s_\theta^2}{c_\theta}
               \left(\frac{\delta g^\prime}{g^\prime}
                    +\frac{1}{2}\delta Z_B\right)\notag\\
      -&\frac{t_3}{2}c_\theta\delta Z_W\left(\frac{1}{4}\delta Z_W
                               -\frac{\delta g}{g}\right)
      +\frac{(Q-t_3)}{2}\frac{s_\theta^2}{c_\theta}
              \delta Z_B\left(\frac{1}{4}\delta Z_B
                             -\frac{\delta g^\prime}{g^\prime}\right)
      \bigg\}\notag\\
  +ig\gamma_\mu\gamma_R\bigg\{&\beta_{R}\delta Z_R
              -Q\frac{s_\theta^2}{c_\theta}
               \left(\frac{\delta g^\prime}{g^\prime}
                    +\frac{1}{2}\delta Z_B\right)
      +\frac{Q}{2}\frac{s_\theta^2}{c_\theta}
              \delta Z_B\left(\frac{1}{4}\delta Z_B
                       -\frac{\delta g^\prime}{g^\prime}\right)\bigg\}
\end{align}
Here $\gamma_L$ and $\gamma_R$ are the left- and right-helicity
projection operators.
It is a simple matter, using eq.(\ref{eq:inversepropform})
and (\ref{eq:photonvertexct}), to show that
the fermion wavefunction counterterms, $\delta Z_L$ and $\delta Z_R$,
cancel between vertex and 2-point counterterms in all Feynman diagrams
of interest here. We will therefore not consider them further.

\subsection{Higgs field lagrangian}

The bare Higgs field after spontaneous symmetry breaking generates
charged and neutral Goldstone scalars $\left(\phi^\pm\right)^0$,
$\phi_Z^0$. Defining the renormalized scalars fields and wavefunction
counterterms via the relation
\begin{equation}
\phi^0=(1+\delta Z_\phi)^{\frac{1}{2}} \phi
\end{equation}
leads to ${\cal O}(\alpha)$ scalar and vector-scalar mixing
counterterms
\begin{eqnarray}
\raisebox{-0.5cm}{
\begin{minipage}[t][0.5cm][c]{2.5cm}
\epsfig{file=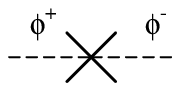,width=2.5cm}
\end{minipage}}
&=&-\delta Z_\phi q^2-\delta\beta\label{eq:phi2ptCT}
\end{eqnarray}
\begin{eqnarray}
\raisebox{-0.5cm}{
\begin{minipage}[t][0.5cm][c]{2.5cm}
\epsfig{file=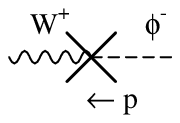,width=2.5cm}
\end{minipage}}
&=&-ip_\mu M_W\left(\delta Z_\phi
                             +\frac{1}{2}\delta Z_W
                             +\frac{\delta g}{g}\right)
\label{eq:WphimixCT}
\end{eqnarray}
here $\delta\beta$ is a quadratically divergent 1-point counterterm.
It will be assumed that it is adjusted to exactly cancel the tadpole
contributions and can therefore be ignored.
A detailed complete renormalization of the Higgs sector along the lines
of ref.\cite{RossTaylor} can be found in ref.\cite{thesis}
where it is shown that
\begin{equation}
\delta Z_\phi^{(1)}
= \frac{\delta M_W^{2(1)}}{M_W^2}-2\frac{\delta g^{(1)}}{g}.
\end{equation}
Note that the fermionic part of this counterterm is finite and in the
$\overline{{\rm MS}}$ renormalization scheme is therefore set to zero.
This is also true of the mixing between neutral scalars and vector
bosons.

\subsection{Gauge-fixing lagrangian}

In the calculations performed here 't Hooft-Feynman, $R_{\xi=1}$, gauge
will be employed for which the gauge-fixing lagrangian is
\begin{equation}
{\cal L}_{{\rm g.f.}}=
   -\frac{1}{2\xi}\left\{
                   2\left|\partial_\mu W_\mu^+-\xi M_W\phi^+\right|^2
        +\left(\partial_\mu Z_\mu-\xi M_Z\phi_Z\right)^2
        +\left(\partial_\mu A_\mu\right)^2
                  \right\}
\label{eq:Lgf}
\end{equation}
with $\xi=1$.
At tree level this has the advantage that
the mixing between vector bosons and Goldstone scalars is canceled
between the vector-scalar interaction lagrangian and the gauge-fixing
lagrangian, ${\cal L}_{{\rm g.f.}}$.
In order to satisfy Ward identities ${\cal L}_{{\rm g.f.}}$ is
constructed from renormalized fields.
Mixing counterterms then reappear in 2-loop diagrams
thereby nullifying an important advantage of $R_\xi$ gauges. Certain
authors \cite{BohmHollSpie} have chosen to replace the renormalized
fields and masses in eq.(\ref{eq:Lgf}) with bare one and satisfy
the Ward identities by renormalizing the gauge parameter, $\xi$. Mixing
counterterms are thus eliminated but no reduction in labour is achieved
because gauge-parameter counterterms now appear and the two approaches
are formally equivalent. The latter, however, goes against the notion
of renormalization as a rescaling of the {\it physical}\/ parameters.
In the present work we follow Ross and Taylor \cite{RossTaylor} leaving
${\cal L}_{{\rm g.f.}}$ unrenormalized.

As is well-known it is only correct to include ${\cal L}_{{\rm g.f.}}$
of eq.(\ref{eq:Lgf}) if the corresponding Faddeev-Popov ghost lagrangian
is included as well. It can be shown that the fermionic part of the
${\cal O}(N_f\alpha)$ 2-point ghost counterterms can only take the form
\begin{eqnarray}
\raisebox{-0.5cm}{
\begin{minipage}[t][0.5cm][c]{2.5cm}
\epsfig{file=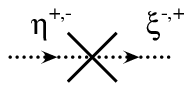,width=2.5cm}
\end{minipage}}
&=&-\delta Z_\xi(q^2+M_W^2)\label{eq:ghost2ptCT}
\end{eqnarray}
where $\delta Z_\xi$ is the ghost wavefunction counterterm.
Note that there is no ghost mass counterterm
{\it per se}\/ which is consistent with the requirement that the ghosts
appear only in closed loops and do not couple directly to fermions.
The presence of the inverse propagator $(q^2+M_W^2)$ in the ghost
counterterm means that contributions from this 2-point counterterm
cancel against diagrams containing ghost vertex counterterms and thus
at ${\cal O}(N_f\alpha^2)$ the ghosts effectively go uncorrected.

\section{Charge Renormalization at ${\cal O}(\alpha)$}
\label{sect:oneloop}

The Thomson scattering amplitude has been calculated by a number of authors
\cite{Sirlin80,Hollik,SirlinMS,ZMass4}. The results of ref.\cite{ZMass4}
are given for a general renormalization scheme assuming only that
renormalization of the bare parameters of the model takes the form given
in eq.(\ref{eq:renorm1}) and eq.(\ref{eq:renorm2})
and the same conventions are adopted here.
For external photon momentum $q^2=0$, the sum of 1-loop photon-fermion
vertex diagrams and external fermion leg corrections is
\begin{equation}
\raisebox{-0.1cm}{
\begin{minipage}[t][0.5cm][c]{2.5cm}
\epsfig{file=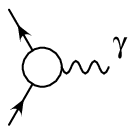,width=2.5cm}
\end{minipage}}
\equiv V_{i\gamma}^{(1b)}(0)\gamma_\mu\gamma_L
=i\frac{g^3 s_\theta}{16\pi^2}2t_3 \gamma_\mu\gamma_L
                    (\pi M_W^2)^{-\epsilon}\Gamma(\epsilon)
\label{eq:oneloopVAi}
\end{equation}
The corresponding vertex and external leg corrections for the $Z^0$-boson
are
\begin{equation}
\raisebox{-0.1cm}{
\begin{minipage}[t][0.5cm][c]{2.5cm}
\epsfig{file=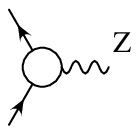,width=2.5cm}
\end{minipage}}
\equiv V_{iZ}^{(1b)}(0)\gamma_\mu\gamma_L
=i\frac{g^3 c_\theta}{16\pi^2}2t_3 \gamma_\mu\gamma_L
                    (\pi M_W^2)^{-\epsilon}\Gamma(\epsilon)
\label{eq:oneloopVZi}
\end{equation}
In both cases care has been taken to use the Feynman rules obtained from the
renormalized lagrangian without applying the relation,
$gs_\theta=g^\prime c_\theta$. This is important when one comes to derive
the 1-loop counterterm insertions at ${\cal O}(N_f\alpha^2)$.

The $Z$-$\gamma$ mixing at $q^2=0$ that contributes to the Thomson
scattering matrix element is given by
\begin{eqnarray}
\raisebox{-0.15cm}{
\begin{minipage}[t][0.5cm][c]{2.75cm}
\epsfig{file=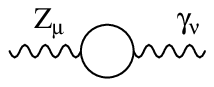,width=2.5cm}
\end{minipage}}
\equiv\delta_{\mu\nu}\Pi_{Z\gamma}^{(1b)}(0)
&=&-2\frac{(g^2+g^{\prime 2})}{16\pi^2}s_\theta c_\theta \delta_{\mu\nu}
   M_W^2(\pi M_W^2)^{-\epsilon}\Gamma(\epsilon)
\label{eq:oneloopZA}\\
&=&-2\frac{g^2}{16\pi^2}s_\theta c_\theta \delta_{\mu\nu}
   M_Z^2(\pi M_W^2)^{-\epsilon}\Gamma(\epsilon)
\end{eqnarray}
where again in eq.(\ref{eq:oneloopZA}) care has been take not to apply
the relation $g s_\theta=g^\prime c_\theta$.
The 1-loop fermionic corrections to the $Z$-$\gamma$ mixing,
$\Pi_{Z\gamma}^{(1f)}(0)$, vanish at $q^2=0$.

The photon self-energy is guaranteed to be purely transverse by
gauge invariance and will be written
\begin{equation}
\raisebox{-0.2cm}{
\begin{minipage}[t][0.5cm][c]{2.5cm}
\epsfig{file=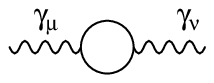,width=2.5cm}
\end{minipage}}
=(q^2\delta_{\mu\nu}-q_\mu q_\nu)\Pi_{\gamma\gamma}^\prime(q^2)
\label{eq:photonse}
\end{equation}
with
\begin{eqnarray}
\Pi_{\gamma\gamma}^{\prime(1b)}(0)&=&
 \left(\frac{g^2 s_\theta^2}{16\pi^2}\right)
 \frac{10}{3}(\pi M_W^2)^{-\epsilon}\Gamma(\epsilon)
+\left(\frac{g^2 s_\theta^2}{16\pi^2}\right)(\pi M_W^2)^{-\epsilon}
                               \epsilon\Gamma(\epsilon)\nonumber\\
&-&\frac{(gs_\theta+g^\prime c_\theta)^2}{16\pi^2}.\frac{1}{12}(\pi
M_W^2)^{-\epsilon}
                               \Gamma(\epsilon)
-\frac{g^{\prime2}c_\theta^2}{16\pi^2}.\frac{1}{3}(\pi M_W^2)^{-\epsilon}
                               \epsilon\Gamma(\epsilon)\\
&=&\left(\frac{g^2 s_\theta^2}{16\pi^2}\right)
   \left(3+\frac{2}{3}\epsilon\right)
       (\pi M_W^2)^{-\epsilon}\Gamma(\epsilon)
\label{eq:photonseb}
\end{eqnarray}
and
\begin{equation}
\Pi_{\gamma\gamma}^{\prime(1f)}(0)=
-\frac{g^{\prime2}c_\theta^2}{16\pi^2}
\frac{4}{3}\pi^{-\epsilon}\Gamma(\epsilon)
\sum_f Q_f^2(m_f^2)^{-\epsilon}
\label{eq:oneloopphotonse}
\end{equation}
where the sum is over internal fermions. When these internal fermions
are quarks eq.(\ref{eq:oneloopphotonse}) cannot be evaluated reliably
because of strong QCD corrections. In that case one writes
\begin{equation}
\Pi_{\gamma\gamma}^{\prime(f)}(0)
=\Re\Pi_{\gamma\gamma}^{\prime(f)}(\hat q^2)
-[\Re\Pi_{\gamma\gamma}^{\prime(f)}(\hat q^2)
    -\Pi_{\gamma\gamma}^{\prime(f)}(0)]
\label{eq:certainidentity}
\end{equation}
with $\hat q^2$ being chosen to be sufficiently large that perturbative
QCD can be used. For $|\hat q^2|\gg m_f^2$
\begin{equation}
\Pi_{\gamma\gamma}^{\prime(1f)}(\hat q^2)=
-\frac{g^{\prime2}c_\theta^2}{16\pi^2}
8(\pi\hat q^2)^{\frac{n}{2}-2}
\frac{\Gamma\left(2-\frac{\displaystyle n}{\displaystyle 2}\right)
            \Gamma\left(\frac{\displaystyle n}{\displaystyle 2}\right)^2}
     {\Gamma(n)}
\label{eq:oneloopphotonsefatq}
\end{equation}
where we have summed over a complete fermion generation.
The quantity in eq.(\ref{eq:certainidentity}) in square brackets is
obtained from the experimentally measured cross-section, $\sigma_h(q^2)$,
for $e^+e^-\rightarrow$hadrons by means of the dispersion relation
\begin{equation}
\Re\Pi_{\gamma\gamma}^{\prime(f)}(\hat q^2)
    -\Pi_{\gamma\gamma}^{\prime(f)}(0)=
-\frac{\hat q^2}{4\pi^2\alpha}\int_{-\infty}^{-4m_\pi^2}
\frac{\sigma_h(q^2)}{q^2-\hat q^2+i\epsilon}dq^2,
\label{eq:dispersion}
\end{equation}
where $m_\pi$ is the mass of the $\pi^0$.
This dispersion integral has been evaluated most precisely for
$\hat q^2=-M_Z^2$
\cite{Zeppenfeld,Jegerlehner,Swartz,DavierHocker,KuhnSteinhauser}.

The 1-loop counterterm contributions to Thomson scattering may be
obtained from eq.s(\ref{eq:Zgamma2ptct}) and (\ref{eq:photonvertexct}).
Combining all the contributions gives the result that to
${\cal O}(\alpha)$
\begin{equation}
\sqrt{4\pi\alpha}=e\left(
         1+\frac{1}{2}\Pi_{\gamma\gamma}^{\prime(1)}(0)
          -\frac{s_\theta}{c_\theta}\frac{\Pi_{Z\gamma}^{(1b)}(0)}{M_Z^2}
               +s_\theta^2\frac{\delta g^{(1)}}{g}
               +c_\theta^2\frac{\delta g^{\prime(1)}}{g^\prime}\right)
\label{eq:oneloopalpha}
\end{equation}
in a general renormalization scheme. The quantity $\alpha$ that appears
on the left-hand side of eq.(\ref{eq:oneloopalpha}) is the experimentally
measured value $\alpha^{-1}=137.036...$, and all parameters on the
right-hand side are the renormalized parameters in the particular
renormalization scheme that has been chosen.

Note that all dependence on the wavefunction renormalization
counterterms, $\delta Z_W$ and $\delta Z_B$ has canceled and, as a
consequence,
one can safely set $\delta Z_W^{(1)}=\delta Z_B^{(1)}=0$ as was done
in ref.s\cite{Sirlin80,ZMass4}. This choice leads to divergent Green
functions that, however, combine to yield finite expressions
for physical quantities such as eq.(\ref{eq:oneloopalpha}). This
cancellation of divergences is a useful and stringent check. It will be
seen, however, that at ${\cal O}(N_f\alpha^2)$ the ${\cal O}(N_f\alpha)$
wavefunction counterterms must be explicitly included in order to obtain
physically correct results. This amounts to imposing the 1-loop Ward
identities by force.

\section{Wave function Renormalization and Ward Identities}
\label{sect:Wardid}

As noted in the foregoing section at 1-loop order,
one has the option of setting
\begin{equation}
\delta Z_W^{(1)}=\delta Z_B^{(1)}=0
\label{eq:dZWdZB0}
\end{equation}
because physical results such as
eq.(\ref{eq:oneloopalpha}) are independent of them. The same cancellation
of the dependence on $\delta Z_W^{(1)}$ and $\delta Z_B^{(1)}$ that occurred
at 1-loop will obviously occur for $\delta Z_W^{(2)}$ and $\delta Z_B^{(2)}$
at 2-loops but then, as will be demonstrated, the condition
(\ref{eq:dZWdZB0}) cannot be maintained and the 1-loop Ward identities
must be imposed explicitly. Actually this feature is seen to be quite
general.
At ${\cal O}(\alpha^n)$ the wavefunction counterterms $\delta Z^{(n)}$
will cancel out in physical expressions but lower order wavefunction
counterterms must be included.

\begin{figure}[t]
\centerline{\epsfig{file=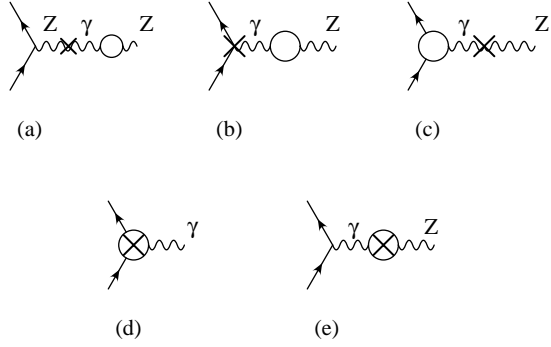}}
\caption{Diagrams containing a 1-loop fermionic counterterm and
         a factor $(\pi M_W^2)^{-\epsilon}\Gamma(\epsilon)$.}
\label{fig:Wardid}
\end{figure}

Consider the diagrams shown in Fig.\ref{fig:Wardid}.
The counterterms denoted by `$\times$' are the fermionic
pieces only and the self-energy and vertex
blobs  represent bosonic radiative corrections. The blobs containing an
`$\times$' denote the bosonic 1-loop diagrams with one-loop fermionic
counterterm insertions. These diagrams are all proportional to the
fermionic part of a 1-loop counterterm and the quantity
$(\pi M_W^2)^{-\epsilon}\Gamma(\epsilon)$ and there are no other diagrams
having this functional dependence. The matrix element for Thomson scattering
is expected to be proportional to the charge, $Q$, of the external fermion
and independent of its weak isospin, $t_3$.
It follows that those parts of the diagrams in Fig.\ref{fig:Wardid}
proportional to $t_3$ must cancel amongst themselves.
By explicit calculation this is found to be
\begin{equation}
ig\left(\frac{g^2}{16\pi^2}\right)2t_3 s_\theta(s_\theta^2-c_\theta^2)
\left\{\frac{\delta g^{(1f)}}{g}
     -\frac{\delta g^{\prime(1f)}}{g^\prime}
+\frac{1}{2}\delta Z_W^{(1f)}-\frac{1}{2}\delta Z_B^{(1f)}\right\}
(\pi M_W^2)^{-\epsilon}\Gamma(\epsilon)
\label{eq:t3ctcancel}
\end{equation}
The 1-loop Ward identities require that
\begin{eqnarray}
\frac{1}{2}\delta Z_B^{(1)}+\frac{\delta g^{\prime(1)}}{g^\prime}&=&0
\label{eq:WardidB}\\
\frac{1}{2}\delta Z_W^{(1f)}+\frac{\delta g^{(1f)}}{g}&=&0
\label{eq:WardidW}
\end{eqnarray}
where eq.(\ref{eq:WardidB}) is true for both fermionic and bosonic
counterterms separately.
Hence (\ref{eq:t3ctcancel}) correctly vanishes provided
$\delta Z_W^{(1f)}$ and $\delta Z_B^{(1f)}$ are included in a manner
consistent with the 1-loop Ward identities. The conditions
(\ref{eq:WardidB}) and (\ref{eq:WardidW}) will therefore be applied
where needed in the following.

It will also be useful to note that in any renomalization scheme
\begin{eqnarray}
\frac{1}{2}\delta Z_W^{(1b)}+\frac{\delta g^{(1b)}}{g}
+2\left(\frac{g^2}{16\pi^2}\right)(\pi M_W^2)^{-\epsilon}\Gamma(\epsilon).
&=&\mbox{finite}
\label{eq:finitecombo1}\\
\delta Z^{(1f)}_\phi=
\frac{\delta M_W^{2(1f)}}{M_W^2}-2\frac{\delta g^{(1f)}}{g}&=&\mbox{finite}
\label{eq:finitecombo2}
\end{eqnarray}
The former vanishes in the on-shell renormalization scheme and the latter
in $\overline{\rm MS}$.

Although rather arduous, we have checked that the sum of terms
proportional to $t_3$ from pure counterterm contributions also vanishes.
This is a useful check of combinatorics and the counterterms
(\ref{eq:Zgamma2ptct}) and (\ref{eq:photonvertexct}).
In this case however the cancellation happens without having to impose
the 1-loop Ward identities explicitly and is valid beyond
${\cal O}(N_f\alpha^2)$ and up to ${\cal O}(\alpha^2)$.

It can also be shown that the ${\cal O}(N_f\alpha^2)$ corrections from
one particle reducible (1PR) proportional to $t_3$ cancel amongst
themselves.

\section{Charge Renormalization at ${\cal O}(N_f\alpha^2)$}

All results given in this section will assume one massless fermion
generation
and in loops in Feynman diagrams will be summed over all fermions in that
generation.

\subsection{One-particle reducible diagrams}

The presence of mixing between the $Z^0$ and the photon greatly
complicates the calculation particularly in the counterterm and
one-particle reducible (1PR) sectors. Baulieu and Coquereaux
\cite{BaulieuCoquereaux} have shown how to treat
$Z$-$\gamma$ mixing to arbitrary order in $\alpha$. Their results can be
applied straightforwardly to neutral current processes such as
$e^+e^-\rightarrow\mu^+\mu^-$ but it is not immediately clear how to treat
the case of Thomson scattering where the photon is external.
In ref.\cite{ZMass6} it was shown how to rearrange the expressions
obtained by Baulieu and Coquereaux in a form that displays the exact
factorization of the residue at the pole of a resonant matrix element
that is known from $S$-matrix theory to occur even in the presence of
mixing.
The same procedure can be used to obtain an exact expression for
the residue at $q^2=0$ for some neutral current process such as
$e^+e^-\rightarrow\mu^+\mu^-$. In that case the initial state residue
factor is found to be
\begin{equation}
\frac{
V_{iZ}(0)\frac{\displaystyle\Pi_{Z\gamma}(0)}
              {\displaystyle M_Z^2-\Pi_{ZZ}(0)}+V_{i\gamma}(0)
     }
     {
\sqrt{
\frac{\displaystyle d}{\displaystyle dq^2}
\left(
q^2-\Pi_{\gamma\gamma}(q^2)
-\frac{\displaystyle \Pi_{\gamma Z}^2(q^2)}
      {\displaystyle M_Z^2-\Pi_{ZZ}(q^2)}
\right)
\Bigg\vert_{q^2=0}
     }
     }
\label{eq:residuefactor}
\end{equation}
which is precisely the Thomson scattering matrix element up to crossings.
Here $V_{iZ}(q^2)$ and $V_{i\gamma}(q^2)$ are the exact, all order,
$e^+e^-Z$ and $e^+e^-\gamma$ vertex corrections respectively.
The self-energy and mixing corrections
$\Pi_{ZZ}(q^2)$, $\Pi_{Z\gamma}(q^2)$ and $\Pi_{\gamma\gamma}(0)$
are also the exact expressions. Expanding the square root generates the
appropriate factors for 1PR diagrams and the correctness of the
procedure is confirmed by the cancellation between the 1PR counterterm
contributions proportional to $t_3$ with those coming from higher-order
counterterms as described in section~\ref{sect:Wardid}.

It was shown in the previous section that the wave function counterterms,
$\delta Z_W^{(1)}$ and $\delta Z_B^{(1)}$ must be included in a manner
consistent with the Ward identities and as stated above
it can also be shown that the
contributions from the 1PR diagrams proportional
to $t_3$ cancel amongst themselves. The calculation can therefore be
organized in such a way that the 1PR diagrams and their
associated counterterms together form a class that is separately
finite and proportional only to the charge, $Q$, of the external fermion.
This also means that there will be a separate cancellation of the
divergences of ${\cal O}(N_f\alpha^2)$ one-particle irreducible
(1PI) diagrams with their associated counterterms. In particular,
it follows that the divergence structure of the 1PI diagrams is not
influenced by the 1PR sector.

\begin{figure}[t]
\centerline{\epsfig{file=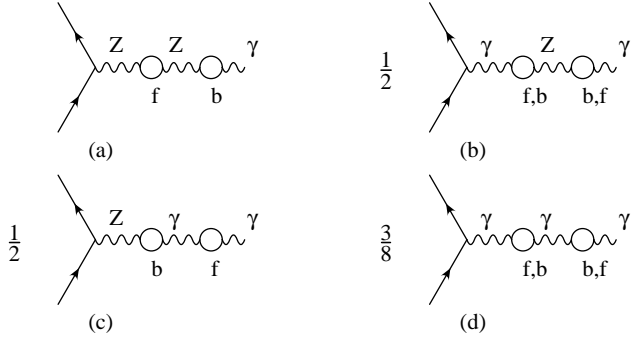}}
\caption{${\cal O}(N_f\alpha^2)$ one-particle reducible (1PR) diagrams.}
\label{fig:1PR}
\end{figure}

The 1PR self-energy
diagrams contributing to the Thomson scattering matrix element
are shown in Fig.\ref{fig:1PR}.
The self-energy blobs are indicated to be
fermionic or bosonic contributions by the `f' or `b' below them.
The associated combinatoric factors are also given.
Note diagrams containing the 1-loop vertex
corrections, (\ref{eq:oneloopVAi}) and (\ref{eq:oneloopVZi}), do not
appear because they are proportional to $t_3$ and have been shown to
cancel as discussed in section~\ref{sect:Wardid}. Discarding all terms
proportional to $t_3$ from the 1PR diagrams their contribution is
\begin{eqnarray}
igs_\theta Q\Pi^{{\rm 1PR}}(0)&=&
igs_\theta^3Q\left(2\frac{\delta g^{(1f)}}{g}
                   -\frac{\delta M_W^{2(1f)}}{M_W^2}\right)
              .\widehat\Pi^{(1b)}_{Z\gamma}(0)
 -igs_\theta^2 c_\theta Q
              \widehat\Pi^{\prime(1f)}_{Z\gamma}(0)
              .\widehat\Pi^{(1b)}_{Z\gamma}(0)
\nonumber\\
 &+&igs_\theta^3\frac{Q}{2}
              \widehat\Pi^{(1b)}_{Z\gamma}(0)
             .\widehat\Pi^{\prime(1f)}_{\gamma\gamma}(0)
 +igs_\theta\frac{3Q}{4}
              \widehat\Pi^{\prime(1f)}_{\gamma\gamma}(0)
             .\widehat\Pi^{\prime(1b)}_{\gamma\gamma}(0)
\label{eq:Pi1PR}
\end{eqnarray}
where
\begin{eqnarray*}
\widehat\Pi^{(1b)}_{Z\gamma}(0)&=&
        2\left(\frac{g^2}{16\pi^2}\right)(\pi M_W^2)^{-\epsilon}
                                         \Gamma(\epsilon)
        +\frac{1}{2}\delta Z_W^{(1b)}+\frac{\delta g^{(1b)}}{g}\\
\widehat\Pi^{\prime(1f)}_{Z\gamma}(0)&=&
                \Pi_{Z\gamma}^{\prime(1f)}(0)
                   +2s_\theta c_\theta\left(
                        \frac{\delta g^{(1f)}}{g}
                       -\frac{\delta g^{\prime(1f)}}{g^\prime}\right)\\
\widehat\Pi^{\prime(1)}_{\gamma\gamma}(0)&=&
        \Pi_{\gamma\gamma}^{\prime(1)}(0)
          +2\left(s_\theta^2\frac{\delta g^{(1)}}{g}
                 +c_\theta^2\frac{\delta g^{\prime(1)}}{g^\prime}
            \right)
\end{eqnarray*}

The eq.s(\ref{eq:WardidB}) and (\ref{eq:WardidW}) have been used along
with the fact $\Pi_{ZZ}^{(1f)}(0)=0$. The quantity
$\Pi_{Z\gamma}^{\prime(1f)}(q^2)$ is the derivative of
$\Pi_{Z\gamma}^{(1f)}(q^2)$ with respect to $q^2$. It will have a
hadronic component for $q^2=0$ that may be obtained using methods
described in ref.\cite{MarcianoSirlin}. This hadronic contribution
is distinct from the one that appears in
$\Pi_{\gamma\gamma}^{\prime(1f)}(0)$ that was discussed in
section~\ref{sect:oneloop}. For the leptons
$\Pi_{\gamma\gamma}^{\prime(1f)}(0)$ may be evaluated perturbatively
from
\begin{equation}
\Pi_{Z\gamma}^{\prime(1f)}(0)=
-\frac{g^2 s_\theta}{16\pi^2}
\frac{4}{3}\pi^{-\epsilon}\Gamma(\epsilon)
\sum_f \left(\frac{\beta_{Lf}+\beta_{Rf}}{2}\right)Q_f
(m_f^2)^{-\epsilon}.
\end{equation}

In the on-shell renormalization scheme all terms vanish identically
because of the definitions of the 1-loop counterterms. This does not
eliminate hadronic contributions, however, because they will reappear
when the counterterms obtained from charge renormalization are used
in other calculations and will give rise to an hadronic uncertainty.

\subsection{Vertex and $Z$-$\gamma$ corrections}

\subsubsection{Diagrams}

\begin{figure}[t]
\centerline{\epsfig{file=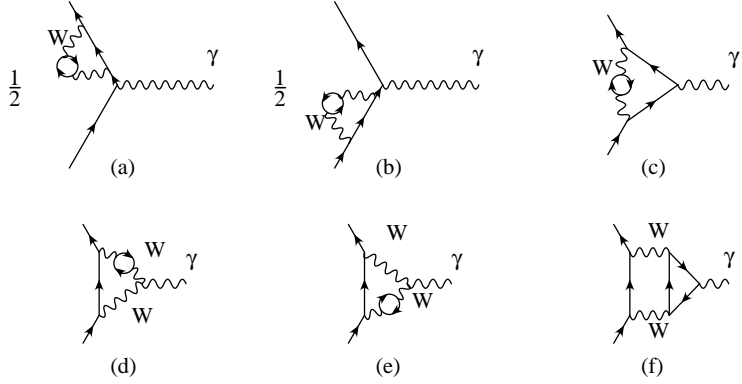}}
\caption{${\cal O}(N_f\alpha^2)$ photon vertex corrections.}
\label{fig:photonvertex}
\end{figure}

Representative topologies for the
${\cal O}(N_f\alpha^2)$ photon vertex diagrams contributing to Thomson
scattering are shown in Fig.\ref{fig:photonvertex}. These may be
calculated
using methods described in ref.\cite{loweng}. Diagrams of the type
Fig.\ref{fig:photonvertex}e--f containing virtual photons or
$Z^0$'s, instead of $W$ bosons, cancel by Ward identities and we have
explicitly checked that this occurs. The sum of all diagrams in
Fig.\ref{fig:photonvertex} is
\begin{eqnarray}
\raisebox{-0.1cm}{
\begin{minipage}[t][0.5cm][c]{2.5cm}
\epsfig{file=vtx1offa.eps,width=2.5cm}
\end{minipage}}
 &\equiv&V_{i\gamma}^{(2)}(0)\gamma_\mu\gamma_L\nonumber\\
 &=&-ig\left(\frac{g^2}{16\pi^2}\right)^2 8 t_3 s_\theta\gamma_\mu\gamma_L
\frac{(\pi M_W^2)^{n-4}}{n}
\Gamma(4-n)\Gamma\left(2-\frac{n}{2}\right)\Gamma\left(\frac{n}{2}\right),
\nonumber\\
\label{eq:Nfalphavertex}
\end{eqnarray}
exactly for all $n$.
\begin{figure}[p]
\centerline{\epsfig{file=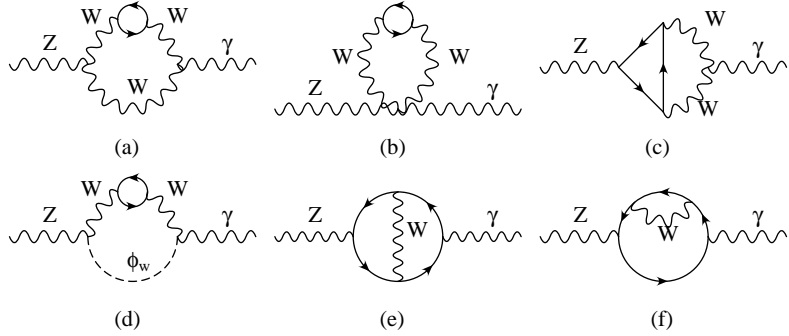}}
\caption{${\cal O}(N_f\alpha^2)$ corrections to the $Z$-$\gamma$ mixing,
         $\Pi^{(2)}_{Z\gamma}(0)$.}
\label{fig:ZAmixing}
\end{figure}

Representative
diagrams contributing to $Z$-$\gamma$ mixing in ${\cal O}(N_f\alpha^2)$
are shown in Fig.\ref{fig:ZAmixing}. Again methods for calculating the
individual diagrams may be found in ref.\cite{loweng}. Upon summing all
diagrams together one obtains the remarkably simple result
\begin{eqnarray}
\raisebox{-0.15cm}{
\begin{minipage}[t][0.5cm][c]{2.75cm}
\epsfig{file=sezoa.eps,width=2.5cm}
\end{minipage}}
 &\equiv&\delta_{\mu\nu}\Pi_{Z\gamma}^{(2b)}(0)\nonumber\\
 &=&\left(\frac{g^2}{16\pi^2}\right)^2 8 s_\theta c_\theta\delta_{\mu\nu}
M_Z^2\frac{(\pi M_W^2)^{n-4}}{n}
\Gamma(4-n)\Gamma\left(2-\frac{n}{2}\right)\Gamma\left(\frac{n}{2}\right).
\nonumber\\
\label{eq:NfZgamma}
\end{eqnarray}

The ${\cal O}(N_f\alpha^2)$ diagrams when added together must form a pure
vector current proportional to the charge, $Q$, of the external fermion
and independent of its weak isospin, $t_3$. Contributions from the photon
self-energy, which will be dealt with later, are obviously of this form and
cancellation of terms proportional to $t_3$ is expected between the vertex
corrections, given above, and the $Z$-$\gamma$ mixing
when it is coupled to the external fermion. This is indeed borne out and
one obtains
\begin{multline}
 V_{i\gamma}^{(2)}(0)\gamma_\mu\gamma_L
+i\frac{g}{c_\theta}\gamma_\mu(t_3\gamma_L-s_\theta^2 Q)
\frac{\Pi_{Z\gamma}^{(2b)}(0)}{M_Z^2}=\\
-ig\left(\frac{g^2}{16\pi^2}\right)^2 8 Q s_\theta^3\gamma_\mu
\frac{(\pi M_W^2)^{n-4}}{n}
\Gamma(4-n)\Gamma\left(2-\frac{n}{2}\right)\Gamma\left(\frac{n}{2}\right).
\label{eq:t3diagcancel}
\end{multline}

\subsubsection{Counterterm Insertions}

The ${\cal O}(N_f\alpha^2)$ corrections coming from the insertion of
1-loop counterterms into 1-loop diagrams may be calculated using the
expressions for the 2-point counterterms given in section~3 and
simplified using the Ward identity (\ref{eq:WardidW}).
Note once again that contributions coming from the first term in the
2-point counterterm for the $W$ boson, (\ref{eq:W2ptct}), cancel against
vertex counterterms. The result is
\begin{multline}
V_{i\gamma}^{(2\delta)}(0)\gamma_\mu\gamma_L
                             =\frac{\delta g^{(1f)}}{g}
                            ig\left(\frac{g^2}{16\pi^2}\right)
                            3t_3s_\theta\gamma_\mu\gamma_L
                            (\pi M_W^2)^{-\epsilon}
                            \Gamma(\epsilon)
\\
                          -\frac{\delta M_W^{2(1f)}}{M_W^2}
                            ig\left(\frac{g^2}{16\pi^2}\right)
                            t_3 s_\theta\gamma_\mu\gamma_L
                            (\pi M_W^2)^{-\epsilon}
                            \epsilon\Gamma(\epsilon)
\label{eq:ViCTresult}
\end{multline}

Similarly the ${\cal O}(N_f\alpha^2)$ contribution from 1-loop counterterm
insertions into the 1-loop $Z$-$\gamma$ mixing is
\begin{multline}
\Pi_{Z\gamma}^{(2\delta)}(0)=\frac{\delta g^{(1f)}}{g}
                            \left(\frac{g^2 s_\theta}{16\pi^2}\right)
                            s_\theta c_\theta M_Z^2
                            (\pi M_W^2)^{-\epsilon}
                            \Gamma(\epsilon)
\\
                          -\delta M_W^{2(1f)}
                            \left(\frac{g^2 s_\theta}{16\pi^2}\right)
                            \frac{s_\theta}{c_\theta}
                            (\pi M_W^2)^{-\epsilon}(2-\epsilon)
                            \Gamma(\epsilon)
\label{eq:ZACTinsert}
\end{multline}
Individual self-energy diagrams contributing to eq.(\ref{eq:oneloopZA})
contain pieces proportional to $g^2s_\theta c_\theta M_W^2$ and
$g^2(s_\theta^3/c_\theta)M_W^2$ and there is a very complex and
intricate interplay between diagrams containing
$W$ boson 2-point counterterms (\ref{eq:W2ptct}),
scalar counterterms (\ref{eq:phi2ptCT})
the vector-scalar mixing counterterms (\ref{eq:WphimixCT})
to produce an overall result
proportional to $g^2(s_\theta/c_\theta)M_W^2$. This, when connected
to the external fermion by a $Z^0$  propagator leads to a result
proportional simply to $g^3s_\theta$ of the same form as
$V_{i\gamma}^{(2\delta)}(0)$ in eq.(\ref{eq:ViCTresult}).

Combining eq.(\ref{eq:ViCTresult}) and eq.(\ref{eq:ZACTinsert}) yields
the total ${\cal O}(N_f\alpha^2)$ contribution counterterm insertions
in the vertex and $Z$-$\gamma$ mixing,
\begin{multline}
 V_{i\gamma}^{(2\delta)}(0)\gamma_\mu\gamma_L
+i\frac{g}{c_\theta}\gamma_\mu(t_3\gamma_L-s_\theta^2 Q)
\frac{\Pi_{Z\gamma}^{(2\delta)}(0)}{M_Z^2}=\\
\left(2\frac{\delta g^{(1f)}}{g}-\frac{\delta M_W^{2(1f)}}{M_W^2}\right)
ig\left(\frac{g^2}{16\pi^2}\right)2t_3s_\theta\gamma_\mu\gamma_L
(\pi M_W^2)^{-\epsilon}\Gamma(\epsilon)\\
-\frac{\delta g^{(1f)}}{g}
ig\left(\frac{g^2}{16\pi^2}\right)Qs_\theta^3\gamma_\mu
(\pi M_W^2)^{-\epsilon}\Gamma(\epsilon)\\
+\frac{\delta M_W^{2(1f)}}{M_W^2}
ig\left(\frac{g^2}{16\pi^2}\right)Qs_\theta^3\gamma_\mu
(\pi M_W^2)^{-\epsilon}(2-\epsilon)\Gamma(\epsilon)
\label{eq:ZAmixandVActinsert}
\end{multline}

Note that the finite terms, proportional to
$t_3 \epsilon\Gamma(\epsilon)$ have canceled. The remaining part
proportional to $t_3$ contributes to eq.(\ref{eq:t3ctcancel})
and therefore cancels with other terms as discussed in
section~\ref{sect:Wardid}. It will be discarded
and only the second and third terms in eq.(\ref{eq:ZAmixandVActinsert}),
proportional to $Q$, will be retained.

\subsubsection{Counterterms}

The expressions for the counterterms given in eq.(\ref{eq:photon2ptct}),
(\ref{eq:Zgamma2ptct}) and (\ref{eq:photonvertexct}) are correct to
${\cal O}(\alpha^2)$ and need to be specialized to ${\cal O}(N_f\alpha^2)$.
Using the Ward identities, (\ref{eq:WardidB}) and (\ref{eq:WardidW}),
the $Z$-$\gamma$ mixing counterterm of eq.(\ref{eq:Zgamma2ptct}) becomes
\begin{eqnarray}
\raisebox{-0.2cm}{
\begin{minipage}[t][0.5cm][c]{2.5cm}
\epsfig{file=sezxa.eps,width=2.5cm}
\end{minipage}}
&=&M_Z^2 s_\theta c_\theta\bigg\{3
     \left(\frac{\delta g^{(1b)}}{g}.\frac{\delta g^{(1f)}}{g}
          -\frac{\delta g^{\prime(1b)}}{g^\prime}
          .\frac{\delta g^{\prime(1f)}}{g^\prime}
     \right)
\nonumber\\ & &\ \ \ \ \ \ \ \ \ \ \ \ \
-\frac{\delta M_W^{2(1f)}}{M_W^2}
     \left(\frac{1}{2}\delta Z_W^{(1b)}+\frac{\delta g^{(1b)}}{g}\right)
     \bigg\}
\end{eqnarray}
and the photon vertex counterterm, eq.(\ref{eq:photonvertexct}), yields
\begin{eqnarray}
\raisebox{-0.2cm}{
\begin{minipage}[t][0.5cm][c]{2.5cm}
\epsfig{file=vtx0cxfa.eps,width=2.5cm}
\end{minipage}}
&=&-igs_\theta t_{3}\gamma_\mu\gamma_L\bigg\{
     3\left(\frac{\delta g^{(1b)}}{g}.\frac{\delta g^{(1f)}}{g}
            -\frac{\delta g^{\prime(1b)}}{g^\prime}
            .\frac{\delta g^{\prime(1f)}}{g^\prime}
       \right)
\nonumber\\ & &\ \ \ \ \ \ \ \ \ \ \ \ \ \ \ \ \ \ \ \
  -2\frac{\delta g^{(1f)}}{g}
     \left(\frac{1}{2}\delta Z_W^{(1b)}+\frac{\delta g^{(1b)}}{g}\right)
     \bigg\}
\nonumber\\ & &\ \ \ \ \ \ \ \ \ \
  -igs_\theta Q\gamma_\mu 3\frac{\delta g^{\prime(1b)}}{g^\prime}
                   .\frac{\delta g^{\prime(1f)}}{g^\prime}
\end{eqnarray}

Their contribution to the Thomson scattering matrix element together is
\begin{multline}
\left(2\frac{\delta g^{(1f)}}{g}-\frac{\delta M_W^{2(1f)}}{M_W^2}\right)
\left(\frac{1}{2}\delta Z_W^{(1b)}+\frac{\delta g^{(1b)}}{g}\right)
ig2t_3s_\theta\gamma_\mu\gamma_L\\
+\frac{\delta M_W^{2(1f)}}{M_W^2}
ig Qs_\theta^3
\left(\frac{1}{2}\delta Z_W^{(1b)}+\frac{\delta g^{(1b)}}{g}\right)
\gamma_\mu\\
-igs_\theta Q\,.\, 3\left(
 s_\theta^2\frac{\delta g^{(1b)}}{g}.\frac{\delta g^{(1f)}}{g}
+c_\theta^2\frac{\delta g^{\prime(1b)}}{g^\prime}.
           \frac{\delta g^{\prime(1f)}}{g^\prime}
\right)\gamma_\mu
\label{eq:ZAmixandVAct}
\end{multline}

As discussed in section~\ref{sect:Wardid} the part proportional to
$t_3$ can be shown to cancel against products of 1-loop counterterms
coming from 1PR diagrams and will therefore be discarded.

It is convenient at this point to define a quantity,
$\widehat\Pi_{Z\gamma}^{(2)}(0)$, obtained by combining the parts
proportional to $Q$ of eq.(\ref{eq:NfZgamma}),
eq.(\ref{eq:ZAmixandVActinsert})
and all but the last term in eq.(\ref{eq:ZAmixandVAct}).
Hence
\begin{eqnarray}
\widehat\Pi_{Z\gamma}^{(2)}(0)&=&
\left(\frac{g^2}{16\pi^2}\right)^2 8 s_\theta c_\theta
M_Z^2\frac{(\pi M_W^2)^{n-4}}{n}
\Gamma(4-n)\Gamma\left(2-\frac{n}{2}\right)\Gamma\left(\frac{n}{2}\right)
\nonumber\\
& &+\frac{\delta g^{(1f)}}{g}
\left(\frac{g^2}{16\pi^2}\right)s_\theta c_\theta M_Z^2
(\pi M_W^2)^{-\epsilon}\Gamma(\epsilon)\nonumber\\
& &-\frac{\delta M_W^{2(1f)}}{M_W^2}
s_\theta c_\theta M_Z^2
\left(\frac{1}{2}\delta Z_W^{(1b)}+\frac{\delta g^{(1b)}}{g}
+2\left(\frac{g^2}{16\pi^2}\right)
(\pi M_W^2)^{-\epsilon}\Gamma(\epsilon)
  \right)\nonumber\\
& &-\frac{\delta M_W^{2(1f)}}{M_W^2}
\left(\frac{g^2}{16\pi^2}\right)s_\theta c_\theta M_Z^2
(\pi M_W^2)^{-\epsilon}(2-\epsilon)\Gamma(\epsilon).
\label{eq:PiZAhat}
\end{eqnarray}
\begin{figure}[p]
\centerline{\epsfig{file=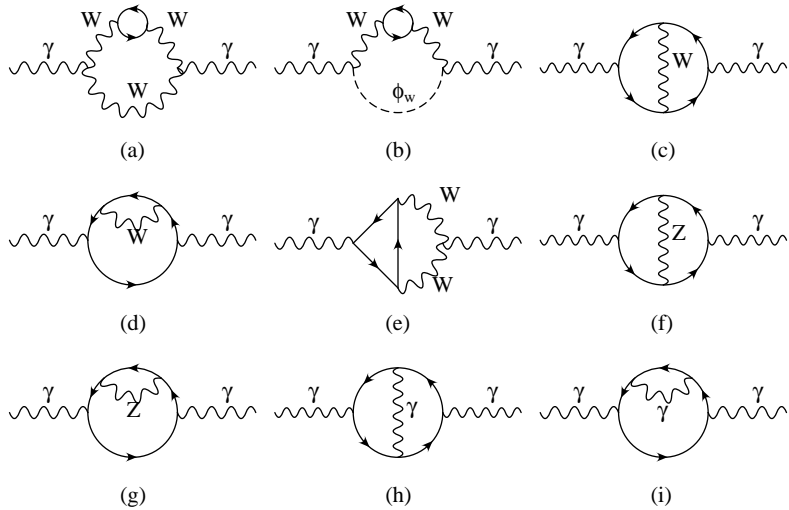}}
\caption{${\cal O}(N_f\alpha^2)$ corrections to the photon self-energy,
         $\Pi^{\prime(2)}_{\gamma\gamma}(0)$.}
\label{fig:photonse}
\end{figure}
\subsection{The Photon Self-Energy}
\label{sect:photonse}

\subsubsection{Diagrams}

Representative topologies for
the diagrams contributing to the photon self-energy,
$\Pi_{\gamma\gamma}^\prime(0)$ of eq.(\ref{eq:photonse})
at ${\cal O}(N_f\alpha^2)$ are shown in Fig.\ref{fig:photonse}.
Calculation of the photon self-energy involves projecting out the
transverse parts of individual diagrams followed by differentiation
with respect to the external momentum squared using techniques
described in ref.\cite{loweng}. Individual diagrams are not separately
transverse but we have checked that the longitudinal part vanishes when all
diagrams are added together. The result is
\begin{eqnarray}
\Pi_{\gamma\gamma}^{\prime(2)}(0)&=&
\left(\frac{g^2 s_\theta}{16\pi^2}\right)^2 \frac{8}{3}
\frac{(n+2)}{n}(\pi M_W^2)^{n-4}
\Gamma(4-n)\Gamma\left(2-\frac{n}{2}\right)
\Gamma\left(\frac{n}{2}-1\right)
\nonumber\\
&+&\left(\frac{g^2 s_\theta}{16\pi^2}\right)^2
\frac{4}{27c_\theta^2}(44 s_\theta^4-27s_\theta^2+9)\nonumber\\
& &\ \ \ \ \ \ \ \ \times
\frac{(n-6)}{n}(\pi M_Z^2)^{n-4}
\Gamma(5-n)\Gamma\left(2-\frac{n}{2}\right)
\Gamma\left(\frac{n}{2}-1\right)
\nonumber\\
&-&\left(\frac{g^2s_\theta}{16\pi^2}\right)^2
\frac{4(n-2)}{3n}\pi^{n-4}(M_W^2)^{\frac{n}{2}-2}
\Gamma\left(3-\frac{n}{2}\right)\Gamma\left(2-\frac{n}{2}\right)
\sum_f Q_f^2(m_f^2)^{\frac{n}{2}-2}
\nonumber\\
&-&\left(\frac{g^2s_\theta}{16\pi^2}\right)^2
\frac{8}{3}\pi^{n-4}(M_W^2)^{\frac{n}{2}-2}
\Gamma\left(2-\frac{n}{2}\right)^2
\sum_f Q_f t_{3f}(m_f^2)^{\frac{n}{2}-2}
\nonumber\\
&-&\left(\frac{g^2s_\theta}{16\pi^2}\right)^2
\frac{16(n-2)}{3n}\pi^{n-4}(M_Z^2)^{\frac{n}{2}-2}
\Gamma\left(3-\frac{n}{2}\right)\Gamma\left(2-\frac{n}{2}\right)
\nonumber\\& &\ \ \ \ \ \ \ \ \times
\sum_f Q_f^2\left(\frac{\beta_{Lf}^2+\beta_{Rf}^2}{2}\right)
       (m_f^2)^{\frac{n}{2}-2}
\nonumber\\
&+&\left(\frac{g^2s_\theta}{16\pi^2}\right)^2
\frac{16n}{3(n-2)}\pi^{n-4}(M_Z^2)^{\frac{n}{2}-2}
\Gamma\left(3-\frac{n}{2}\right)\Gamma\left(2-\frac{n}{2}\right)
\nonumber\\& &\ \ \ \ \ \ \ \ \times
\sum_f Q_f^2\beta_{Lf}\beta_{Rf}(m_f^2)^{\frac{n}{2}-2}
\nonumber\\
&+&\left(\frac{g^2s_\theta}{16\pi^2}\right)^2
\frac{4s_\theta^2}{3}
\frac{(5n^2-33n+34)}{n(n-5)}\pi^{n-4}
\Gamma\left(3-\frac{n}{2}\right)\Gamma\left(2-\frac{n}{2}\right)
\sum_f Q_f^4 (m_f^2)^{n-4}
\nonumber\\
\label{eq:twoloopphotonse}
\end{eqnarray}
Contributions that are suppressed by factors $m_f^2/M_W^2$ relative to the
leading terms have been dropped.
The first term on the right hand side of eq.(\ref{eq:twoloopphotonse})
comes from diagrams Fig.\ref{fig:photonse}a--e
that contain an internal $W$ boson, the
second comes from diagrams Fig.\ref{fig:photonse}f\&g
containing an internal $Z^0$. The last term in
eq.(\ref{eq:twoloopphotonse}) comes from diagrams
Fig.\ref{fig:photonse}h\&i that are pure QED in nature.
Contributions for which the fermion mass can be safely
set to zero without affecting the final result were be obtained
using the methods
described in ref.\cite{loweng}. The terms in which the fermion mass
appears are obtained using the asymptotic expansion of ref.\cite{DavyTausk}.
It should be noted that setting $m_f=0$ in
Fig.\ref{fig:photonse}c--g does not immediately cause
obvious problems in the computation because the diagram still contains
one non-vanishing scale. A certain amount of care is thus required
to identify situations in which the fermion mass cannot be discarded.
In the case of the photon self-energy, the need to include such terms
is indicated by divergences proportional to the $\ln m_f^2$
in the counterterm insertion diagrams.
All diagrams that yield $m_f$-dependent terms can be split in two by a
cut through two internal fermion propagators. The contributions are
therefore precisely those that are accessed via dispersion relations
(\ref{eq:dispersion}).

As it stands the eq.(\ref{eq:twoloopphotonse}) for
$\Pi_{\gamma\gamma}^{\prime(2)}(0)$ contains a divergence, in its fourth
term, with a coefficient that depends on $\ln m_f^2$. It will be
seen in the next section that this is canceled by a counterterm insertion.
There remain finite terms depending on the fermion mass that must be
treated using dispersion relations when the internal fermions are quarks.

\subsubsection{Counterterm Insertions}

The ${\cal O}(N_f\alpha^2)$ corrections coming from 1-loop counterterm
insertions into 1-loop diagrams for the transverse part of the
photon self-energy may be calculated to be
\begin{eqnarray}
\Pi_{\gamma\gamma}^{\prime(2\delta)}(0)&=&
             2\frac{\delta g^{(1f)}}{g}
              \left(\frac{g^2s_\theta^2}{16\pi^2}\right)
           (\pi M_W^2)^{-\epsilon}\Gamma(\epsilon)\nonumber\\
& &+\frac{\delta M_W^{2(1f)}}{M_W^2}
    \left(\frac{g^2s_\theta^2}{16\pi^2}\right)
           \left(\frac{2}{3}\epsilon-5\right)
    (\pi M_W^2)^{-\epsilon}\epsilon\Gamma(\epsilon)\nonumber\\
& &-\left(\frac{\delta g^{(1b)}}{g}+\frac{1}{2}\delta Z_W^{(1b)}\right)
   \left(\frac{g^2s_\theta^2}{16\pi^2}\right)
   \frac{4}{3}\pi^{-\epsilon}\Gamma(\epsilon)
   \sum_f Q_f t_{3f} (m_f^2)^{-\epsilon}\nonumber\\
& &+\left(\frac{g^2s_\theta^2}{16\pi^2}\right)
   \frac{8}{3}\pi^{-\epsilon}\epsilon\Gamma(\epsilon)
   \sum_f Q_f^2\frac{\delta m_f^{(1b)}}{m_f} (m_f)^{-\epsilon}
\label{eq:photontwoloopctinsert}
\end{eqnarray}

It was checked that the longitudinal form factor vanishes which involves,
once again, an intricate interplay between diagrams containing
$W$-boson 2-point counterterms (\ref{eq:W2ptct}),
scalar counterterms (\ref{eq:phi2ptCT})
and vector-scalar mixing counterterms (\ref{eq:WphimixCT}).

By virtue of (\ref{eq:finitecombo1}),
the second term on the right-hand side of
eq.(\ref{eq:photontwoloopctinsert}) cancels the divergence in
the fourth term of eq.(\ref{eq:twoloopphotonse})
that depends on $\ln m_f^2$. It may also be seen by substituting
the expression for $\delta m_f$ in eq.(\ref{eq:fermionmassct})
into eq.(\ref{eq:photontwoloopctinsert})
that the remaining terms that depend
on $m_f$ in $\Pi_{\gamma\gamma}^{\prime(2)}(0)$
are rendered finite by the
counterterm insertions with the exception of the last.
Moreover, in the on-shell renormalization
scheme, there is an exact cancellation and these finite terms are
eliminated completely. At 1-loop the bosonic and fermionic sectors of the
theory are renormalized independently of one another.
It is obviously a great convenience
and simplification here to demand that the internal fermions are
renormalized
in the on-shell scheme. This will be done in the following but no such
constraint will be imposed on the bosonic sector.

Combining eq.(\ref{eq:twoloopphotonse}) and
eq.(\ref{eq:photontwoloopctinsert}) we define a new quantity,
$\widehat\Pi_{\gamma\gamma}^{\prime(2)}(0)=
\Pi_{\gamma\gamma}^{\prime(2)}(0)+
\Pi_{\gamma\gamma}^{\prime(2\delta)}(0)$, given by
\begin{eqnarray}
\widehat\Pi_{\gamma\gamma}^{\prime(2)}(0)&=&
\left(\frac{g^2 s_\theta}{16\pi^2}\right)^2 \frac{8}{3}
\frac{(n+2)}{n}(\pi M_W^2)^{n-4}
\Gamma(4-n)\Gamma\left(2-\frac{n}{2}\right)
\Gamma\left(\frac{n}{2}-1\right)
\nonumber\\
&+&\left(\frac{g^2 s_\theta}{16\pi^2}\right)^2
\frac{4}{27c_\theta^2}(44 s_\theta^4-27s_\theta^2+9)\nonumber\\
& &\ \ \ \ \ \ \ \ \times
\frac{(n-6)}{n}(\pi M_Z^2)^{n-4}
\Gamma(5-n)\Gamma\left(2-\frac{n}{2}\right)
\Gamma\left(\frac{n}{2}-1\right)
\nonumber\\
&+&2\left(\frac{g^2s_\theta^2}{16\pi^2}\right)
\frac{\delta g^{(1f)}}{g}
(\pi M_W^2)^{\frac{n}{2}-2}\Gamma\left(2-\frac{n}{2}\right)
\nonumber\\
&-&\left(\frac{g^2s_\theta^2}{16\pi^2}\right)
\frac{\delta M_W^{2(1f)}}{M_W^2}
\left(\frac{n+11}{3}\right)
(\pi M_W^2)^{\frac{n}{2}-2}\Gamma\left(3-\frac{n}{2}\right)
\nonumber\\
&-&\left(\frac{g^2 s_\theta^2}{16\pi^2}\right)
\frac{4}{3}
\left(\frac{\delta g^{(1b)}}{g}+\frac{1}{2}\delta Z_W^{(1b)}
+2\left(\frac{g^2}{16\pi^2}\right)
(\pi M_W^2)^{\frac{n}{2}-2}
\Gamma\left(2-\frac{n}{2}\right)
\right)
\nonumber\\& &\ \ \ \ \ \ \ \ \times
\pi^{\frac{n}{2}-2}
\Gamma\left(2-\frac{n}{2}\right)
\sum_f Q_f t_{3f}(m_f^2)^{\frac{n}{2}-2}
\nonumber\\
&+&\left(\frac{g^2s_\theta^2}{16\pi^2}\right)^2
\frac{4(n^3-12n^2+41n-34)}{n(n-3)(n-5)}\pi^{n-4}
\Gamma\left(3-\frac{n}{2}\right)\Gamma\left(2-\frac{n}{2}\right)
\nonumber\\& &\ \ \ \ \ \ \ \ \times
\sum_f Q_f^4 (m_f^2)^{n-4}
\label{eq:PiAAhat}
\end{eqnarray}

Of the two remaining terms in eq.(\ref{eq:PiAAhat}) that depend on
$m_f$, the first corresponds to weak corrections to photon-fermion
vertex that have their origin in the diagram of
Fig.\ref{fig:photonse}e.
The second comes from the pure QED diagrams,
Fig.\ref{fig:photonse}h\&i, and their associated
fermion mass counterterms.
When expanded about $n=4$ its leading logarithms reproduce the well-known
result of Jost and Luttinger \cite{JostLuttinger}.
Both sets can be treated via the dispersion
relation trick, eq.(\ref{eq:certainidentity}). This requires that the
diagrams be evaluated at some high $\hat q^2$.
Fig.\ref{fig:photonse}e at high
$\hat q^2$ contains two unrelated scales and cannot be treated by the
techniques used so far.
In principle it can be obtained in closed analytic form
from results given by Scharf and Tausk \cite{ScharfTausk} but it is
neither compact nor illuminating and probably best obtained numerically.

The pure QED diagrams of Fig.\ref{fig:photonse}h\&i are
exactly calculable for $|\hat q^2|\gg m_f^2$. The result is
\begin{multline}
\Pi_{\gamma\gamma}^{\prime(2{\rm QED})}(\hat q^2)=
-\sum_f\left(\frac{g^2}{16\pi^2}\right)^2
Q_f^4 s_\theta^4(\pi\hat q^2)^{n-4}\\
\times
\Bigg\{8(n^2-7n+16)
     \frac{
          \Gamma\left(2-\frac{\displaystyle n}{\displaystyle 2}\right)^2
          \Gamma\left(\frac{\displaystyle n}{\displaystyle 2}\right)^3
          \Gamma\left(\frac{\displaystyle n}{\displaystyle 2}-2\right)
          }
          {\Gamma(n)\Gamma(n-1)}\\
+24\frac{(n^2-4n+8)}{(n-1)(n-4)}.
     \frac{
          \Gamma(4-n)
          \Gamma\left(\frac{\displaystyle n}{\displaystyle 2}\right)^2
          \Gamma\left(\frac{\displaystyle n}{\displaystyle 2}-2\right)
          }
          {\Gamma\left(\frac{\displaystyle 3n}{\displaystyle 2}-2\right)}
\Bigg\}
\label{eq:highqphotonse}
\end{multline}
Broadhurst {\it et al.}\ \cite{Broadhurst} have given an expression
for the subtracted photon vacuum polarization at general $q^2$.
The high-energy limit can be obtained by applying analytic continuation
relations for the hypergeometric functions, ${}_2F_1$ and ${}_3F_2$
that appear in their result.

Despite appearances, the expression on the right hand side of
eq.(\ref{eq:highqphotonse}) has only a simple pole with a constant
coefficient at $n=4$ that can be canceled by local counterterms.
The leading logarithmic expressions can be found in
ref.\cite[section 8-4-4]{ItzyksonZuber} where the authors invite the
``foolhardy reader'' to check that the finite parts are transverse.
Here we have gone further and demonstrated this property in the exact
result.

It can be checked that the divergent part of (\ref{eq:highqphotonse})
is identical to that of the last term in eq.(\ref{eq:PiAAhat})
so that that the difference
$\Re\Pi_{\gamma\gamma}^{\prime(f)}(\hat q^2)
    -\Pi_{\gamma\gamma}^{\prime(f)}(0)$ is finite.
This provides yet another useful check of various aspects of the
calculation.

\subsection{Charge renormalization in a general scheme}

All the ingredients are now in place to write down the complete set
of ${\cal O}(N_f\alpha^2)$ corrections to the Thomson scattering
matrix element and from it obtain an expression, extending
eq.(\ref{eq:oneloopalpha}), for the physical
electromagnetic charge of a fermion in terms of the renormalized
parameters of the theory.

Gathering all the parts together gives our main result
\begin{align}
\sqrt{4\pi\alpha}=e\bigg\{
         1&+\frac{1}{2}\Pi_{\gamma\gamma}^{\prime(1)}(0)
          -\frac{s_\theta}{c_\theta}\frac{\Pi_{Z\gamma}^{(1b)}(0)}{M_Z^2}
               +s_\theta^2\frac{\delta g^{(1)}}{g}
               +c_\theta^2\frac{\delta g^{\prime(1)}}{g^\prime}
\nonumber\\
         &+\frac{1}{2}\widehat\Pi_{\gamma\gamma}^{\prime(2)}(0)
          -\frac{s_\theta}{c_\theta}
           \frac{\widehat\Pi_{Z\gamma}^{(2)}(0)}{M_Z^2}
           +\Pi^{{\rm 1PR}}(0)
               +s_\theta^2\frac{\delta g^{(2)}}{g}
               +c_\theta^2\frac{\delta g^{\prime(2)}}{g^\prime}
\nonumber\\
          &-3\left(s_\theta^2\frac{\delta g^{(1b)}}{g}.
                             \frac{\delta g^{(1f)}}{g}
                  +c_\theta^2\frac{\delta g^{\prime(1b)}}{g^\prime}.
                             \frac{\delta g^{\prime(1f)}}{g^\prime}
             \right)\bigg\}
\end{align}
where $\widehat\Pi_{\gamma\gamma}^{\prime(2)}(0)$
is given in eq.(\ref{eq:PiAAhat}),
$\widehat\Pi_{Z\gamma}^{(2)}(0)$ in eq.(\ref{eq:PiZAhat}) and
$\Pi^{{\rm 1PR}}(0)$ in eq.(\ref{eq:Pi1PR}).

\section{${\cal O}(N_f\alpha^2)$ wavefunction counterterms}

Up to this point we have been pursuing the expression for the physical
matrix element for Thomson scattering and, to that end, terms that do not
contribute to the final result have often been discarded. Although the final
result does not depend on the ${\cal O}(N_f\alpha^2)$ wavefunction
renormalization counterterms, $\delta Z_W^{(2)}$ and $\delta Z_B^{(2)}$,
the Green's functions that have been calculated in the foregoing allow
relations to be determined between them and $\delta g^{(2)}$ and
$\delta g^{\prime(2)}$. In all cases
the leading divergences of these counterterms, i.e.\ those corresponding
to a double pole at $n=4$, are independent of renormalization scheme
but subleading and finite parts will depend on which renormalization
scheme has been chosen.

When the ${\cal O}(N_f\alpha^2)$ diagrams contributing to $Z$-$\gamma$
mixing (\ref{eq:NfZgamma}) are combined with the counterterm insertions
(\ref{eq:ZACTinsert}) and the counterterms (\ref{eq:Zgamma2ptct})
the result must be finite in any scheme.
One thus obtains
\begin{align}
&\left(\frac{g^2}{16\pi^2}\right)^2
 8\frac{(\pi M_W^2)^{n-4}}{n}
\Gamma(4-n)\Gamma\left(2-\frac{n}{2}\right)\Gamma\left(\frac{n}{2}\right)
+\frac{\delta g^{(1f)}}{g}
\left(\frac{g^2}{16\pi^2}\right)
(\pi M_W^2)^{-\epsilon}\Gamma(\epsilon)
\nonumber\\
&+\frac{\delta M_W^{2(1f)}}{M_W^2}
\left(\frac{g^2}{16\pi^2}\right)
(\pi M_W^2)^{-\epsilon}\epsilon\Gamma(\epsilon)
\nonumber\\
&
-\frac{\delta M_W^{2(1f)}}{M_W^2}
     \left(\frac{1}{2}\delta Z_W^{(1b)}+\frac{\delta g^{(1b)}}{g}
     +2\left(\frac{g^2}{16\pi^2}\right)
       (\pi M_W^2)^{-\epsilon}\Gamma(\epsilon)
     \right)
 \\
&  +3\left(\frac{\delta g^{(1b)}}{g}.\frac{\delta g^{(1f)}}{g}
          -\frac{\delta g^{\prime(1b)}}{g^\prime}
          .\frac{\delta g^{\prime(1f)}}{g^\prime}
     \right)
-\frac{\delta g^{(2)}}{g}+\frac{\delta g^{\prime(2)}}{g^\prime}
-\frac{1}{2}\delta Z_W^{(2)}+\frac{1}{2}\delta Z_B^{(2)}
=\mbox{finite}\nonumber
\end{align}

Turning to the ${\cal O}(N_f\alpha^2)$ corrections to the photon
vertex it similarly follows that when the contributions from diagrams,
(\ref{eq:Nfalphavertex}), counterterm insertions, (\ref{eq:ViCTresult}),
and pure counterterms (\ref{eq:photonvertexct}) are added together
the result is finite.
The part of the vertex proportional to $t_3$ yields the condition
\begin{align}
&\left(\frac{g^2}{16\pi^2}\right)^2
 8\frac{(\pi M_W^2)^{n-4}}{n}
\Gamma(4-n)\Gamma\left(2-\frac{n}{2}\right)\Gamma\left(\frac{n}{2}\right)
+\frac{\delta g^{(1f)}}{g}
\left(\frac{g^2}{16\pi^2}\right)
(\pi M_W^2)^{-\epsilon}\Gamma(\epsilon)
\nonumber\\
&+\frac{\delta M_W^{2(1f)}}{M_W^2}
\left(\frac{g^2}{16\pi^2}\right)
(\pi M_W^2)^{-\epsilon}\epsilon\Gamma(\epsilon)
\nonumber\\
&
-2\frac{\delta g^{(1f)}}{g}
     \left(\frac{1}{2}\delta Z_W^{(1b)}+\frac{\delta g^{(1b)}}{g}
     +2\left(\frac{g^2}{16\pi^2}\right)
       (\pi M_W^2)^{-\epsilon}\Gamma(\epsilon)
     \right)
\label{eq:t3term}\\
&  +3\left(\frac{\delta g^{(1b)}}{g}.\frac{\delta g^{(1f)}}{g}
          -\frac{\delta g^{\prime(1b)}}{g^\prime}
          .\frac{\delta g^{\prime(1f)}}{g^\prime}
     \right)
-\frac{\delta g^{(2)}}{g}+\frac{\delta g^{\prime(2)}}{g^\prime}
-\frac{1}{2}\delta Z_W^{(2)}+\frac{1}{2}\delta Z_B^{(2)}
=\mbox{finite}\nonumber
\end{align}

This expression differs from the previous one but is consistent with it
because of the finiteness of the combinations (\ref{eq:finitecombo1})
and (\ref{eq:finitecombo2}). This consistency is a stringent check
of very many aspects of the calculation.

The part of the the photon vertex proportional to $Q$ leads to the condition
\begin{equation}
\frac{\delta g^{\prime(2)}}{g^\prime}+\frac{1}{2}\delta Z_B^{(2)}
   -3\left(\frac{\delta g^{\prime(1b)}}{g^\prime}
          .\frac{\delta g^{\prime(1f)}}{g^\prime}
     \right)=\mbox{finite}.
\label{eq:Qterm}
\end{equation}

Obviously eq.s(\ref{eq:t3term}) and (\ref{eq:Qterm}) can be used
to obtain a condition for the combination of counterterms
\[
\frac{\delta g^{(2)}}{g}+\frac{1}{2}\delta Z_W^{(2)}
+2\frac{\delta g^{(1f)}}{g}
     \left(\frac{1}{2}\delta Z_W^{(1b)}+\frac{\delta g^{(1b)}}{g}\right)
-3\frac{\delta g^{(1b)}}{g}.\frac{\delta g^{(1f)}}{g}
\]
which is the ${\cal O}(N_f\alpha^2)$ $W$-fermion coupling counterterm
and occurs, for example, in the calculation of corrections to the muon
lifetime. Indeed this fact has been exploited as an extremely
stringent cross check of both the results of this paper and of the
calculation ${\cal O}(N_f\alpha^2)$ corrections to the muon lifetime
\cite{muondecay}.

The corrections to photon self-energy calculated in
section~\ref{sect:photonse} yield an independent condition on the
the ${\cal O}(N_f\alpha^2)$ counterterms
\begin{equation}
s_\theta^2\delta Z_W^{(2)}+c_\theta^2\delta Z_B^{(2)}+
\widehat\Pi_{\gamma\gamma}^{\prime(2)}(0)=\mbox{finite}
\end{equation}
where $\widehat\Pi_{\gamma\gamma}^{\prime(2)}(0)$
is given in eq.(\ref{eq:PiAAhat}).

\section{Summary}

The complete ${\cal O}(N_f\alpha^2)$ renormalization of the electromagnetic
charge in the Standard Model using a general renormalization scheme has
been presented. This represents the first practical calculation in which
the full structure of the 2-loop renormalization has been confronted and
lays the groundwork for future calculations of this type. The
results have already been exploited in the calculation of the
${\cal O}(N_f\alpha^2)$ corrections to the muon lifetime.

The $Z$-$\gamma$ mixing adds considerably to the
overall complexity and number of Feynman diagrams that must be considered.
Contributions coming from the insertion of 1-loop
counterterms in 1-loop diagrams constitute a significant
portion of the calculation due partly to the appearance at this order
of counterterms that mix vector bosons with scalars.

All integrals were performed exactly in dimensional regularization without
expanding in $\epsilon=2-n/2$ yet in many cases they produced new and
remarkably simple expressions that display the full analytic structure
of the results.

The r\^ole of wavefunction counterterms was investigated. It was found
that the ${\cal O}(\alpha)$ wavefunction counterterms,
$\delta Z^{(1)}$, had to be included in a manner consistent with Ward
identities but that the ${\cal O}(N_f\alpha^2)$ counterterms,
$\delta Z^{(2)}$, could be neglected since they cancel in the final
physical result. The price for this is that one must deal with divergent
Greens functions in intermediate steps.

A large number of internal consistency checks were performed in the
course of the calculation in order to ensure the correctness of the
results presented here.


\begin{thebibliography}{99}

\bibitem{Zeppenfeld} A. D. Martin and D. Zeppenfeld,
       {\sl Phys.\ Lett.}\ {\bf B 345} (1995) 558.

\bibitem{Jegerlehner} S. Eidelman and F. Jegerlehner,
        {\sl Z. Phys.}\ {\bf C 76} (1995) 585.

\bibitem{Swartz} M. L. Swartz,
       {\sl Phys.\ Rev.}\ {\bf D 53} (1996) 5268.

\bibitem{DavierHocker} M. Davier and A. H\"ocker, hep-ph/9711308.

\bibitem{KuhnSteinhauser} J. H. K\"uhn and M. Steinhauser, hep-ph/9802241.

\bibitem{StuartHadron} R. G. Stuart
       {\sl Phys.\ Rev.}\ {\bf D 52} (1995) 1655.

\bibitem{Barbieri1} R. Barbieri {\it et al.},
        {\sl Phys.\ Lett.}\ {\bf B 288} (1992) 95;
        errata {\it ibid}\/ {\bf B 312} (1993) 511.

\bibitem{Barbieri2} R. Barbieri {\it et al.},
        {\sl Nucl.\ Phys.}\ {\bf B 409} (1993) 105.

\bibitem{Fleischer1} J. Fleischer, O. V. Tarasov and F. Jegerlehner,
        {\sl Phys.\ Lett.}\ {\bf B 319} (1993) 249.

\bibitem{Fleischer2} J. Fleischer, O. V. Tarasov and F. Jegerlehner,
        {\sl Phys.\ Rev.}\ {\bf D 51} (1995) 3820.

\bibitem{DegrGambVici} G. Degrassi, P. Gambino and A. Vicini,
        {\sl Phys.\ Lett.}\ {\bf B 383} (1996) 219.

\bibitem{CKM1} A. Czarnecki, B. Krause and W. J. Marciano,
        {\sl Phys.\ Rev.}\ {\bf D 52} (1995) 2619.

\bibitem{CKM2} A. Czarnecki, B. Krause and W. J. Marciano,
        {\sl Phys.\ Rev. Lett.}\ {\bf 76} (1996) 3267.

\bibitem{RossTaylor} D. A. Ross and J. C. Taylor,
                    {\sl Nucl. Phys.}\ {\bf B 51} (1973) 125.

\bibitem{Aokietal} K. I. Aoki {\it et al}, or,
            {\sl Suppl. Progr. Theor. Phys.}\ {\bf 73} (1982) 1.

\bibitem{BaulieuCoquereaux} L. Baulieu and R. Coquereaux,
                      {\sl Ann. Phys.}\ {\bf 140}\ (1982)\ 163.

\bibitem{thesis} R. G. Stuart, D. Phil. thesis, University of Oxford;
        Rutherford-Appleton Laboratory Report RAL-T008, (1985)

\bibitem{BohmHollSpie} M. B\"ohm, W. Hollik and H. Spiesberger,
    {\sl Fortschr.\ Phys.} {\bf 34} (1986) 687.

\bibitem{Sirlin80} A. Sirlin, {\sl Phys.\ Rev.}\ {\bf D 22} (1980) 971.

\bibitem{Hollik} W. F. L. Hollik,
    {\sl Fortschr.\ Phys.} {\bf 38} (1990) 165.

\bibitem{SirlinMS} A. Sirlin, {\sl Phys.\ Lett.}\ {\bf B 232} (1990) 537.

\bibitem{ZMass4} R. G. Stuart, {\sl Phys.\ Lett.}\ {\bf B 272} (1991) 353.

\bibitem{ZMass6} R. G. Stuart, {\sl Phys. Rev. Lett.}\ {\bf 70} (1993) 3193.

\bibitem{MarcianoSirlin} W. J. Marciano and A. Sirlin,
                {\sl Phys.\ Rev.}\ {\bf D 22} (1980) 2695;
                (E) {\bf D 31} (1985) 231.

\bibitem{loweng}
    R. Akhoury, P. Malde and R. G. Stuart, hep-ph/9707520.

\bibitem{DavyTausk} A. I. Davydychev and J. B. Tausk,
                    {\sl Nucl. Phys.}\ {\bf B 397} (1993) 123.

\bibitem{JostLuttinger} R. Jost and J. M. Luttinger,
     {\sl Helv. Phys. Acta} {\bf 23} (1950) 201.

\bibitem{ScharfTausk} R. Scharf and J. B. Tausk,
                    {\sl Nucl. Phys.}\ {\bf B 412} (1994) 523.

\bibitem{Broadhurst} D. J. Broadhurst, J. Fleischer and O. V. Tarasov,
        {\sl Z.\ Phys.}\ {\bf C 60} (1993) 287.

\bibitem{ItzyksonZuber} C. Itzykson and J.-B. Zuber,
                 {\sl Quantum Field Theory}, McGraw-Hill (1980).

\bibitem{muondecay} P. Malde and R. G. Stuart,
        {\it in preparation}.



\end{thebibliography}
\end{document}